\documentclass[journal]{IEEEtran}

\usepackage{booktabs} % for top, middle and bottomline
\usepackage{multirow} % multi column and row spanning
\usepackage{multicol}
\usepackage{balance}
\usepackage{cite}
\usepackage[switch]{lineno} 
\usepackage{enumerate}
\usepackage{dsfont}
\usepackage{caption}
\usepackage{fancyhdr}
\makeatletter

\newsavebox{\@linebox}
\savebox{\@linebox}[3em][t]{\parbox[t]{3em}{%
    \@tempcnta\@ne\relax
    \loop{\underline{\scriptsize\the\@tempcnta}}\\
    \advance\@tempcnta by \@ne\ifnum\@tempcnta<60\repeat}}
\newif\ifdraft
\ifdraft
\makeatother
\fi
\usepackage{hyperref}
\hypersetup{
  bookmarks=true,         % show bookmarks bar?
  unicode=false,          % non-Latin characters in Acrobat’s bookmarks
  pdftoolbar=true,        % show Acrobat’s toolbar?
  pdfmenubar=true,        % show Acrobat’s menu?
  pdffitwindow=false,     % window fit to page when opened
  pdfstartview={FitH},    % fits the width of the page to the window
  pdftitle={My title},    % title
  pdfauthor={Jun Xing Chin},     % author
  pdfsubject={Subject},   % subject of the document
  pdfcreator={Creator},   % creator of the document
  pdfproducer={Producer}, % producer of the document
  pdfkeywords={keyword1} {key2} {key3}, % list of keywords
  pdfnewwindow=true,      % links in new window
  colorlinks=true,       % false: boxed links; true: colored links
  linkcolor=blue,          % color of internal links (change box color with linkbordercolor)
  citecolor=blue,        % color of links to bibliography
  filecolor=blue,      % color of file links
  urlcolor=black           % color of external links
}

\ifCLASSINFOpdf
\usepackage[pdftex]{graphicx}
\graphicspath{{./}}
\DeclareGraphicsExtensions{.pdf,.jpeg,.eps,.png}
\else
\fi

\usepackage[cmex10]{amsmath}
\usepackage{amssymb}

\usepackage[caption=false,font=footnotesize]{subfig}
\DeclareMathOperator*{\argmin}{arg\,min}
\newcommand{\floor}[1]{\lfloor #1 \rfloor}
\newcommand{\ccdot}{\, \cdot \,}
\newcommand{\tilX}{\tilde{X}}
\newcommand{\tilY}{\tilde{Y}}
\newcommand{\tilI}{\tilde{I}}

\begin{document}

\title{~\\Privacy-Protecting Energy Management Unit through Model-Distribution Predictive Control}
\renewcommand{\theenumi}{\alph{enumi}}
\author{
  
  \IEEEauthorblockN{Jun-Xing~Chin,~\IEEEmembership{Student Member,~IEEE,}
  Tomas~Tinoco~De~Rubira,
  and~Gabriela~Hug,~\IEEEmembership{Senior Member,~IEEE}}%
  
  \thanks{This work was supported in part by the \emph{Swiss National Science Foundation} for the COPES project under the CHIST-ERA Resilient Trustworthy Cyber-Physical Systems (RTCPS) initiative, and in part by the \emph{ETH Zurich Postdoctoral Fellowship FEL-11 15-1}.}%
  
  \thanks{J.X. Chin, T. Tinoco De Rubira, and G. Hug are with the Power Systems Laboratory, ETH Zurich, 8092 Zurich, Switzerland. Email: \{chin $|$ tomast $|$ hug\}@eeh.ee.ethz.ch.}%
}

\fancypagestyle{firstpage}{% Page style for first page
	\fancyhf{}% Clear header/footer
	\renewcommand\headrule{}
	\setlength{\voffset}{-0.4cm}
	\fancyhead[C]{\scriptsize This is the author's version of an article that has been published in this journal. Changes were made to this version by the publisher prior to publication.\\
		The final version of record is available at~ \href{http://dx.doi.org/10.1109/TSG.2017.2703158}{\color{blue}{http://dx.doi.org/10.1109/TSG.2017.2703158}}\\~\\}
	\fancyhead[L]{\footnotesize IEEE TRANSACTIONS ON SMART GRID}
	\fancyfoot[CO]{\scriptsize~\\~\\~\\Copyright (c) 2017 IEEE. Personal use is permitted. For any other purposes, permission must be obtained from the IEEE by emailing \href{mailto:pubs-permissions@ieee.org}{pubs-permissions@ieee.org.}}
}

\IEEEoverridecommandlockouts

\pagestyle{fancy}
\renewcommand\headrule{}
	\setlength{\voffset}{-0.4cm}
	\fancyhead[C]{\scriptsize This is the author's version of an article that has been published in this journal. Changes were made to this version by the publisher prior to publication.\\
		The final version of record is available at~ \href{http://dx.doi.org/10.1109/TSG.2017.2703158}{\color{blue}{http://dx.doi.org/10.1109/TSG.2017.2703158}}\\~\\}
	\fancyhead[L]{\footnotesize IEEE TRANSACTIONS ON SMART GRID}
	\fancyfoot[CO]{~\\\scriptsize Copyright (c) 2017 IEEE. Personal use is permitted. For any other purposes, permission must be obtained from the IEEE by emailing \href{mailto:pubs-permissions@ieee.org}{pubs-permissions@ieee.org.}}

\maketitle
\thispagestyle{firstpage}

%\IEEEpeerreviewmaketitle

%%%%%%%%%%%%%%%%
%%%%%%%%%%%%%%%%
\begin{abstract}
The roll-out of smart meters in electricity networks introduces risks for consumer privacy due to increased measurement frequency and granularity. Through various Non-Intrusive Load Monitoring techniques, consumer behavior may be inferred from their metering data. In this paper, we propose an energy management method that reduces energy cost and protects privacy through the minimization of information leakage. The method is based on a Model Predictive Controller that utilizes energy storage and local generation, and that predicts the effects of its actions on the statistics of the actual energy consumption of a consumer and that seen by the grid. Computationally, the method requires solving a Mixed-Integer Quadratic Program of manageable size whenever new meter readings are available. We simulate the controller on generated residential load profiles with different privacy costs in a two-tier time-of-use energy pricing environment. Results show that information leakage is effectively reduced at the expense of increased energy cost. The results also show that with the proposed controller the consumer load profile seen by the grid resembles a mixture between that obtained with Non-Intrusive Load Leveling and Lazy Stepping.
\end{abstract}


%%%%%%%%%%%%%%%%%%%%%%
%%%%%%%%%%%%%%%%%%%%%%
\section{Introduction}

% Motivation
Globally, traditional electromechanical electricity meters are being replaced by Smart Meters (SMs) as part of efforts to modernize the grid in order to better manage electricity generation and distribution. In Europe, this is mandated by the European Commission under its Third Energy Package, requiring member states to roll-out SMs where cost-benefit analyses are positive \cite{EuropeanUnion2009a}. However, with their ability to provide real-time information on consumer demand, SMs also raise concerns regarding consumer privacy \cite{Quinn2009, McDaniel2009}. By measuring energy demand at much higher frequencies than traditional meters, detailed consumer load profiles can be extracted from SM data. Through Non-Intrusive Load Monitoring (NILM) techniques, which were first studied by Hart in $1989$ \cite{Hart1989} and further developed through the years \cite{Altrabalsi2014, Guanchen2015}, individual appliance usage, and ultimately consumer lifestyle patterns, preferences, and occupancy profiles can be inferred \cite{Molina-Markham2010, Lisovich2010, Greveler2012}. Such data may not only be accessible to Utility Providers (UPs) but also to malicious third parties, as metering infrastructures are vulnerable to attacks that lead to information leakage \cite{McLaughlin2010}. Moreover, if one also considers UPs being untrustworthy parties, privacy concerns stemming from SM roll-outs become much more severe, and can even lead to backlashes against their installation \cite{Hoenkamp2011}.

% Literature review
To counter this, various works have been done in order to protect consumer privacy. While some focus on cyber measures, such as encryption, adding noise to achieve differential privacy \cite{Acs2011}, and consumer aggregation, others address the problem through physical means, \emph{e.g.}, load control \cite{Dong2015} and battery load hiding (BLH) \cite{Kalogridis2010, McLaughlin2011}. BLH techniques mask consumer load by charging and discharging a battery to alter the load profile captured by SMs. For example, in \cite{Kalogridis2010}, the authors use a best-effort algorithm to hide consumer load change, whilst \cite{McLaughlin2011} details a load-leveling method that switches its energy consumption based on the battery's state-of-charge. The authors in \cite{Yang2012}, on the other hand, introduce a stepping algorithm that improves upon the performance of best-effort and load-leveling algorithms. While privacy loss metrics, \emph{e.g.}, Mutual Information and Fischer Information, have been used to assess BLH schemes such as in \cite{Yang2012}, the schemes themselves have been primarily heuristics-driven. Cyber-physical methods proposed for minimizing directly a specific privacy loss measure such as mutual information have so far been mainly theoretical \cite{Giaconi2016, Li2015}.

% Literature review
In general, BLH methods proposed in the literature for privacy protection have not directly considered energy cost in the control policies. Any cost savings obtained have been a by-product of the privacy protection scheme. In these methods, the level of privacy protection is closely linked to storage capacity \cite{Yang2012, Tan2013}, but privacy protection alone is unlikely to justify the cost of investment on high-capacity storage devices, which according to estimates by \cite{Agency2015} remains high. Utilizing energy storage for minimizing energy cost in addition to providing privacy protection is critical for justifying this investment. Methods that focus on minimizing energy cost alone have been the subject of many works. Examples of recent work include \cite{Khakimova2015, Haiming2015}, where the authors utilize batteries in a Model Predictive Control (MPC) scheme to minimize energy costs in a varying-price environment by charging during low-price periods to compensate consumer load during high-price periods.

% Our work
In this paper, we propose a new scheme based on MPC that combines energy cost minimization with privacy protection that directly minimizes mutual information. The proposed scheme introduces binary variables in the MPC optimization subproblems for counting predicted observations and estimating the joint statistics of consumer load and the net load seen by the grid (grid load). This allows the controller to predict the effects of its actions on the mutual information between these two load profiles. Computationally, the proposed scheme requires solving Mixed-Integer Quadratic Programs (MIQPs) of manageable size whenever new meter readings are available. 

% Structure
The rest of this paper is structured as follows: Section \ref{GenProb} describes the problem considered. Section \ref{meth} presents our proposed solution approach. Section \ref{imp} details implementation and numerical experiments. Lastly, Section \ref{Conclusion} summarizes this work and provides an outlook on future research.
  
%%%%%%%%%%%%%%%%%%%%%%%%%%%%%
%%%%%%%%%%%%%%%%%%%%%%%%%%%%%
\section{Problem Description}
\label{GenProb}

\begin{figure}
\centering
\includegraphics[width=\columnwidth]{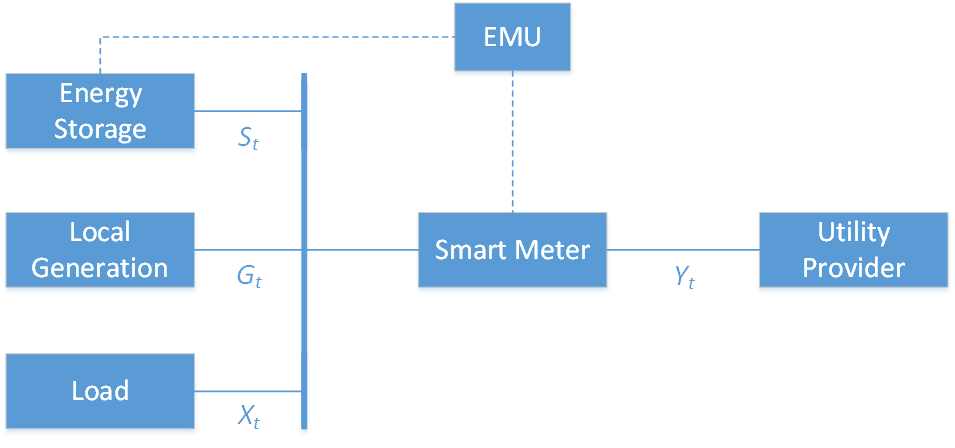}
\vspace{-0.6cm}
\caption{Energy System Model.}
\label{SysMod}
\vspace{-0.5cm}
\end{figure}

% System
We consider the problem of minimizing energy cost and reducing information leakage about the load profile of an energy consumer. The system considered consists of an Energy Management Unit (EMU) that uses energy storage, local generation, and energy available from the grid to supply the consumer load in a cost-effective and privacy-concerned manner. This is illustrated in Fig. \ref{SysMod}. Mathematically, the system is represented by the discrete-time random process
\begin{equation*}
\big\{(S_t, G_t, C_t, X_t, Y_t) \ | \ t \in \mathds{Z}_+ \big\},
\end{equation*}
where $S_t$ is the energy supplied to the energy storage, $G_t$ is the energy supplied by the local generation, $C_t$ is the price of energy, $X_t$ is the consumer energy demand, and $Y_t$ is the energy demand seen by the grid during time interval $t$, \emph{i.e.}, 
\begin{equation}
Y_t = X_t+S_t-G_t. \label{powerflow}
\end{equation}
This implies that no energy wastage is allowed. The realizations of these random variables are denoted by $s_t$, $g_t$, $c_t$, $x_t$, and $y_t$, respectively, for each time $t$. 

% System model

% Details
The battery charging energy $S_t$ is constrained by
\begin{equation}
S^{\min} \leq S_t \leq S^{\max} \label{powerlimits}
\end{equation}
during each time interval $t$, where $-S^{\min} \ge 0$ is the maximum discharging energy and $S^{\max} \ge 0$ is the maximum charging energy at a constant power over a single time interval. These constraints are given by the power rating of the device. It is assumed that the charge and discharge efficiencies are equal, and we denote these with $\alpha \in (0,1)$. The state of charge of the energy storage device at the beginning of time interval $t$ is denoted by $E_t$ and satisfies
\begin{equation} \label{dynamics}
E_{t+1} = E_t + \alpha \mathds{1}\{ S_t \ge 0\} S_t + \alpha^{-1} \mathds{1}\{ S_t < 0 \} S_t,
\end{equation} 
for each $t$, where $\mathds{1}\{A\} = 1$ when $A$ is true and $0$ otherwise. It is also assumed for simplicity that the energy storage device is not used to provide ancillary services to the grid, allowing the full use of its capacity, and as such is constrained by 
\begin{equation}\label{batterycap}
0 \leq E_t \leq E^{\max}, 
\end{equation}
where $E^{\max}$ is the maximum energy storage capacity of the device. Additionally, it is assumed in our system that the local generation is deterministic in nature and that the EMU has perfect prior knowledge regarding its output. The future consumer loads $X_t$ as well as the energy prices $C_t$ are also assumed to be known to the EMU. In practice, only noisy predictions would be available but the incorporation of such uncertainties is part of a future work. With regards to the grid, we assume that no energy feed-in to the grid is allowed and that the energy supplied is bounded by $Y^{\max}$, the maximum connection capacity available on the distribution feeder cables (or circuit breakers), \emph{i.e.}, 
\begin{equation}
0 \leq  Y_t \leq Y^{\max}. \label{ybounds}
\end{equation}

% Energy cost and privacy
The energy cost at each time interval $t$ is given by $C_t Y_t$. On the other hand, loss of consumer privacy is measured by the information leakage rate of consumer load given grid load, which following \cite{Galka2005} is given by
\begin{align*}
I\big(X^{\floor{t}};Y^{\floor{t}}\big) := 
& \int_X \int_Y p_{X,Y}(x_0,y_0,\ldots,x_t,y_t) \, \times \\
& \log \frac{p_{X,Y}(x_0,y_0,\ldots,x_t,y_t)}{p_X(x_0,\ldots,x_t)p_Y(y_0,\ldots,y_t)} \mathop{dy} \mathop{dx} .
\end{align*}
Here, $X^{\floor{t}} := (X_0,\ldots,X_t)$, $Y^{\floor{t}} := (Y_0,\ldots,Y_t)$, $p_{X,Y}$, $p_X$ and $p_Y$ denote the probability density functions of $\big(X^{\floor{t}},Y^{\floor{t}}\big)$, $X^{\floor{t}}$ and $Y^{\floor{t}}$, respectively, and $\log$ denotes the base-$2$ logarithm. Quantifying the cost of privacy loss through a non-negative constant $\mu$ with units of Rp-per-bit (100 Rp = 1 CHF), we seek to find a causal and implementable control policy $U \in \mathcal{U}$ that minimizes the total time-average expected cost. Such a policy determines the charge and discharge of the energy storage device during each time interval $t$ based on the observation history up to that time, \emph{i.e.},  
\begin{equation*}
S_{t} = U(X^{\floor{t}},C^{\floor{t}},G^{\floor{t}},Y^{\floor{t-1}},S^{\floor{t-1}}), 
\end{equation*}
where $C^{\floor{t}}$, $G^{\floor{t}}$ and $S^{\floor{t}}$ are defined in analogous ways to $X^{\floor{t}}$ and $Y^{\floor{t}}$. Mathematically, the problem can be posed as finding a policy $U^*$ such that  
\begin{equation} 
U^{*} = \argmin_{U \in \mathcal{U}} f(U), \label{problem}
\end{equation} 
where 
\begin{equation*}
f(U) := \lim_{\bar{T} \rightarrow \infty} \frac{1}{\bar{T}+1} \left\{ \sum_{\tau=0}^{\bar{T}} \mathds{E}[C_{\tau} Y_{\tau}] + \mu I\big(X^{\floor{\bar{T}}};Y^{\floor{\bar{T}}}\big)\right\}
\end{equation*}
and $\mathds{E}[\ccdot]$ denotes expectation.

%%%%%%%%%%%%%%%%%%%%%%%%%%%%%%%%%%%%%%%%%%%%%%%
%%%%%%%%%%%%%%%%%%%%%%%%%%%%%%%%%%%%%%%%%%%%%%%
\section{Model-Distribution Predictive Control}
\label{meth}

% Overview
We propose an approach based on MPC to find a control policy that is close to $U^*$. At time $t$, the controller (EMU) observes the realizations $x_t$, $g_t$, $c_t$, and $e_t$ (energy storage state of charge), determines actions $s_t,\ldots,s_{t+T}$ for a prediction horizon of length $T$, executes action $s_t$ to achieve a desirable grid load $y_t$, and repeats the process in a receding horizon manner.

% Privacy proxy
Since the exact evaluation of the information leakage rate $I(X^{\floor{t}};Y^{\floor{t}})$ is not possible without knowing the probability density functions $p_{X,Y}$ and $p_Y$, which depend on the control actions, we propose using an approximation. This is done by first assuming that for $\tau$ near time $t$, $(X_{\tau},Y_{\tau})$ are independent identically distributed (i.i.d.) samples of a pair of random variables $(\tilX_t,\tilY_t)$. More specifically, this is assumed during the time window $\{t+T-N+1,\ldots,t+T\}$, which has length $N \gg T$ and covers both the recent past and the entire prediction horizon with respect to time $t$. While time independence does not hold in reality, we make this strong assumption as a first step in making the problem tractable. Time correlation and other properties will be studied in future work. Letting $X^{[t]} := (X_{t-M+1},\ldots,X_{t+T})$ and $Y^{[t]} := (Y_{t-M+1},\ldots,Y_{t+T})$, where $M := N-T$, and approximating the average information leakage per time interval using
\begin{equation*}
    \frac{I(X^{\floor{t+T}};Y^{\floor{t+T}})}{t+T+1} \approx \frac{I(X^{[t]};Y^{[t]})}{N},
\end{equation*}
the locally i.i.d. assumption then gives that 
\begin{align}
\frac{I(X^{\floor{t+T}};Y^{\floor{t+T}})}{t+T+1} 
&\approx \frac{I(X^{[t]};Y^{[t]})}{N} \notag \\ 
& = \!\! \sum_{\tau=t-M+1}^{t+T} \!\!\! \frac{I(X_{\tau};Y_{\tau})}{N} \notag \\ 
& = \!\! \sum_{\tau=t-M+1}^{t+T} \!\!\! \frac{I(\tilde{X}_t;\tilde{Y}_t)}{N} \notag \\
& = I(\tilde{X}_t,\tilde{Y}_t), \label{Iapprox1}
\end{align}
where first equality stems from time independence, and the second equality from the random variables being identically distributed.

\begin{figure}
\centering
\vspace{-0.3cm}
\includegraphics[width=\columnwidth]{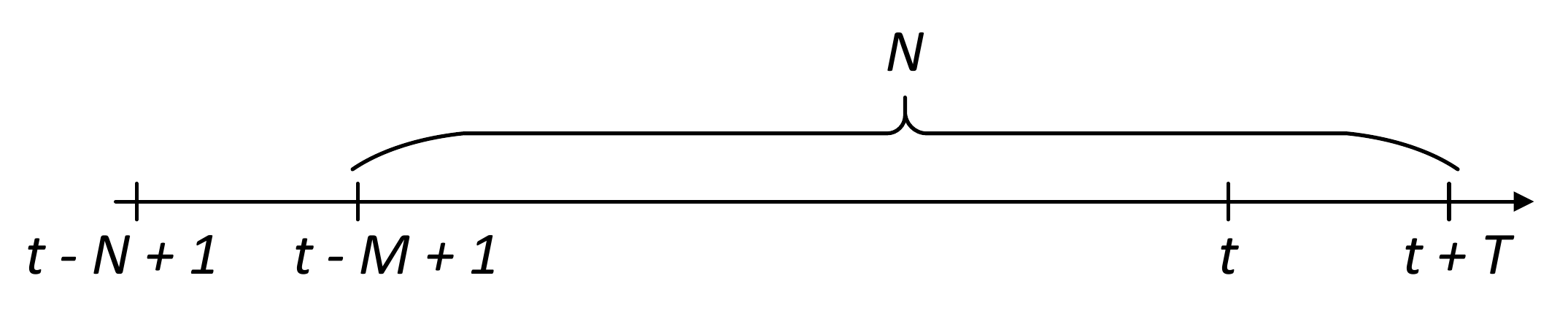}
\vspace{-0.6cm}
\caption{Counting window for probability estimation at time $t$.}
\label{SamplingTime}
\vspace{-0.3cm}
\end{figure}

Furthermore, for the purpose of approximating mutual information only, we assume that the random variables $\tilX_t$ and $\tilY_t$ can take on only a finite number of energy levels that are evenly spaced by $2\Delta$, \emph{i.e.}, $\tilX_t \in \{x^{1},x^{2},...,x^{m}\}$ and $\tilY_t \in \{y^{1},y^{2},...,y^{n}\}$, where $x^{1}-\Delta = 0$, $x^{m} + \Delta = X^{\max}$,  $y^{1}-\Delta = 0$, and $y^{n} + \Delta = Y^{\max}$. The value $X^{\max}$ can be computed from historical records. From this assumption, it follows that
\begin{align}
I(\tilX_t;\tilY_t) 
& = \sum_{i=1}^m\sum_{j=1}^n p_{\tilX,\tilY}(x^i,y^j) \log \frac{p_{\tilX,\tilY}(x^i,y^j)}{p_{\tilX}(x^i) p_{\tilY}(y^j)} \notag \\ 
& = \sum_{i=1}^m\sum_{j=1}^n p_{\tilX,\tilY}(x^i,y^j) \log \frac{p_{\tilY|\tilX}(y^j|x^i)}{p_{\tilY}(y^j)}, \label{Iapprox1_dis}
\end{align} 
where $p_{\tilX,\tilY}$, $p_{\tilX}$, $p_{\tilY}$ and $p_{\tilY|\tilX}$ denote the probability mass functions of $\big(\tilX_t,\tilY_t\big)$, $\tilX_t$, $\tilY_t$ and $\tilY_t$ given $\tilX_t$, respectively. Since $(X_{\tau},Y_{\tau})$ are assumed to be i.i.d samples of $(\tilX_t,\tilY_t)$ for $\tau \in \{t+T-N+1,\ldots,t+T\}$, the statistics of these random variables are approximated by the relative frequency of events during this time window, as illustrated in Fig. \ref{SamplingTime}. More specifically,
\begin{align*}
p_{\tilX,\tilY}(x^i,y^j) & \approx \frac{ \sum_{\tau=t-M+1}^{t+T} \mathds{1}\{(x_{\tau},y_{\tau})=(x^i,y^j)\} + \varepsilon}{N+mn\varepsilon} \\ 
p_{\tilX}(x^i) & \approx \frac{ \sum_{\tau=t-M+1}^{t+T} \mathds{1}\{x_{\tau}=x^i\} + n\varepsilon}{N+mn\varepsilon} \\
p_{\tilY}(y^j) & \approx \frac{ \sum_{\tau=t-M+1}^{t+T} \mathds{1}\{y_{\tau}=y^j\} + m\varepsilon}{N+mn\varepsilon},
\end{align*}
where $\varepsilon > 0$. The addition of the positive scalar $\varepsilon$ in the probability estimates corresponds to additive smoothing \cite{Manning2008}. This is done to avoid probability estimates of zero for events for which there are no observations during the time window. This probability estimation strategy, albeit simplistic, constitutes an adequate choice for a first step towards exploring the proposed methodology of including in MPC the ability to estimate the statistical effects of the control actions in a tractable manner. Including more sophisticated probability estimation techniques into this methodology is an interesting subject for future work.

% Counting window

% Binary variables
Since at time $t$ the counting window covers the recent past and the prediction horizon, the probability estimates can be separated into parts that are constants and parts that depend on the controller's actions during the prediction horizon. More specifically, they can be expressed as
\begin{align}
p_{\tilX,\tilY}(x^i,y^j) & \approx a^{ij}_t + \frac{1}{N_{\varepsilon}}\sum_{\tau=t}^{t+T} z^{ij}_{\tau} \label{pyx1} \\ 
p_{\tilY}(y^j) & \approx b^j_t + \frac{1}{N_{\varepsilon}}\sum_{\tau=t}^{t+T} \sum_{k=1}^m z^{kj}_{\tau}, \label{py1}
\end{align}
where $N_{\varepsilon} := N+mn\varepsilon$, $a^{ij}_t$ and $b^j_t$ are constants (at time $t$) defined by
\begin{align*}
a^{ij}_t & := \frac{\sum_{\tau=t-M+1}^{t-1} \mathds{1}\{(x_{\tau},y_{\tau})=(x^i,y^j)\} + \varepsilon}{N_{\varepsilon}} \\
b^j_t & := \frac{\sum_{\tau=t-M+1}^{t-1} \mathds{1}\{y_{\tau}=y^j\} + m\varepsilon}{N_{\varepsilon}},
\end{align*}
and $z^{ij}_{\tau} := \mathds{1}\{(x_{\tau},y_{\tau})=(x^i,y^j)\}$ for $\tau \in \{t,\ldots,t+T\}$, $i \in \{1,\ldots,m\}$ and $j \in \{1,\ldots,n\}$ are binary variables that depend on the controller's actions and forecasts during the prediction horizon. Letting $c^i_t := p_{\tilX}(x^i)$, which is a constant at time $t$ due to the perfect prediction assumption on consumer load, it follows from \eqref{pyx1} and \eqref{py1} that
\begin{equation*}
p_{\tilY|\tilX}(y^j|x^i) \approx \frac{a^{ij}_t}{c^i_t} + \frac{1}{c^i_tN_{\varepsilon}}\sum_{\tau=t}^{t+T} z^{ij}_{\tau}. \\ 
\end{equation*}
From these expressions for probability estimates and \eqref{Iapprox1_dis}, we then have the approximation
\begin{alignat}{2}
I(\tilX_t;\tilY_t)
&\approx    &&\,\,\bar{I}(\tilX_t;\tilY_t) \nonumber \\
&:=         &&\, \sum_{i=1}^m\sum_{j=1}^n \left( a^{ij}_t + \frac{1}{N_{\varepsilon}}\sum_{\tau=t}^{t+T} z^{ij}_{\tau} \right) \times \nonumber\\
&           &&\left\{ \log \left( a^{ij}_t + \frac{1}{N_{\varepsilon}}\sum_{\tau=t}^{t+T} z^{ij}_{\tau} \right) \right. - \nonumber\\ 
&           &&\left. \log \left( b^j_t + \frac{1}{N_{\varepsilon}}\sum_{\tau=t}^{t+T} \sum_{k=1}^m z^{kj}_{\tau} \right) - \log c^i_t \right\},\label{nonlinear}
\end{alignat}
which is a fairly complicated function of the binary variables $z^{ij}_{\tau}$. Since $N \gg T$, this approximation can be simplified by replacing the logarithms with their first-order Taylor expansions around the constant part of their arguments, \emph{i.e.},
\begin{alignat*}{2}
\log \left( a^{ij}_t + \frac{1}{N_{\varepsilon}}\sum_{\tau=t}^{t+T} z^{ij}_{\tau} \right) 
&& \approx & \log a^{ij}_t \ + \\ 
&&         & \log^{\prime} \big(a^{ij}_t\big) \left(\frac{1}{N_{\varepsilon}}\sum_{\tau=t}^{t+T} z^{ij}_{\tau}\right) \\
\log \left( b^j_t + \frac{1}{N_{\varepsilon}}\sum_{\tau=t}^{t+T} \sum_{k=1}^m z^{kj}_{\tau} \right)
&& \approx & \log b^j_t \ + \\
&&         & \log^{\prime} \big(b^j_t\big) \left(\frac{1}{N_{\varepsilon}}\sum_{\tau=t}^{t+T} \sum_{k=1}^m z^{kj}_{\tau} \right).   
\end{alignat*}
Using these Taylor expansions and letting $\nu := 1/\log_e 2$, we obtain the approximation
\begin{alignat}{2} 
I(\tilX_t;\tilY_t)
&\approx   &&\,\,\tilde{I}(\tilX_t;\tilY_t) \notag \\
&:=      &&\,\sum_{i=1}^m\sum_{j=1}^n \left( a^{ij}_t + \frac{1}{N_{\varepsilon}}\sum_{\tau=t}^{t+T} z^{ij}_{\tau} \right) \times \notag \\
&         &&\left\{ \log \frac{a^{ij}_t}{b^j_t c^i_t} + \frac{\nu}{a^{ij}_t N_{\varepsilon}}\sum_{\tau=t}^{t+T} z^{ij}_{\tau} \right. - \notag \\
&         && \quad\quad\quad\quad\quad \ \left. \frac{\nu}{b^j_t N_{\varepsilon}}\sum_{\tau=t}^{t+T} \sum_{k=1}^m z^{kj}_{\tau} \right\}, \label{linlog}
\end{alignat}
which is now a quadratic function of the binary variables $z^{ij}_{\tau}$.  

% Slope bounds
An important observation is that the curvature of the logarithm functions grows rapidly as zero is approached. Since the constants $a^{ij}_t$ and $b^j_t$ can be potentially very small for some $i$ and $j$ due to the absence of observations of the associated events in the time window $\{t-M+1,\ldots,t-1\}$ and a poor choice of $\varepsilon$ for additive smoothing, the Taylor expansions of the logarithms can be very poor. Hence, we propose choosing a bound $\rho$ for $\log^{\prime} \big(a^{ij}_t\big)$ and $\log^{\prime} \big(b^j_t\big)$ that ensures that the logarithms are always expanded sufficiently away from zero. Since 
\begin{equation*}
\log^{\prime} \big(a^{ij}_t\big) = \frac{\nu}{a^{ij}_t} \le \frac{N + mn\varepsilon}{\varepsilon} \nu
\end{equation*}
and 
\begin{equation*}
\log^{\prime} \big(b^j_t\big) = \frac{\nu}{b^j_t} \le \frac{\nu}{a^{ij}_t} = \log^{\prime} \big(a^{ij}_t\big),
\end{equation*}
it follows that setting $\varepsilon \ge N/(\rho \nu^{-1}-mn)$ ensures that both $\log^{\prime} \big(a^{ij}_t\big)$ and $\log^{\prime} \big(b^j_t\big)$ are always bounded by $\rho$. 

% MPC optimization subproblem
Using \eqref{Iapprox1} and the definition of $\tilI(\tilX_t,\tilY_t)$, the ideal objective function for the MPC optimization problem that exactly captures energy cost minimization and privacy protection is approximated as follows:
\begin{align*}
\frac{1}{T+1} \sum_{\tau=t}^{t+T} c_{\tau} y_{\tau} + \frac{1}{T+t+1} \mu I\big(X^{\floor{t+T}};Y^{\floor{t+T}}\big) \\
\approx \frac{1}{T+1} \sum_{\tau=t}^{t+T} c_{\tau} y_{\tau} + \mu \tilI\big(\tilX_t;\tilY_t\big).
\end{align*}
Hence, the proposed MPC scheme consists of solving at each time $t$ the optimization problem
\begin{alignat}{2}
& \underset{s,e,y,z,w}{\mbox{minimize}} \quad && \frac{1}{T+1} \sum_{\tau=t}^{t+T} c_{\tau} y_{\tau} + \mu \Phi(z) \label{MIQP} \\
& \mbox{subject to} \quad && (s,e,y,z,w) \in \mathcal{F}_t, \notag
\end{alignat}
where $\Phi(z) := \tilI\big(\tilX_t;\tilY_t\big)$ measures privacy loss,  
\begin{equation*}
s := \{s_{\tau}\}_{\tau=t}^{t+T}, \quad e := \{e_{\tau}\}_{\tau=t}^{t+T}, \quad y := \{y_{\tau}\}_{\tau=t}^{t+T}
\end{equation*}
are continuous optimization variables, $z$ is the collection of binary optimization variables defined by
\begin{align*}
z := \left\{ z^{ij}_{\tau} \, \Bigg| \, \right.
& \tau \in \{t,\ldots,t+T\}, \\ 
& j \in \{1,\ldots,n\}, \\ 
& \left. i = \argmin_{k \in \{1,\ldots,m\}} \|x_{\tau}-x^k\|_2 \right\}, 
\end{align*}
and $w := \{w_{\tau}\}^{t+T}_{\tau=t}$ are binary variables used to represent the indicator functions in \eqref{dynamics}. Note that not all binary variables used in \eqref{nonlinear} are treated as optimization variables in \eqref{MIQP}. This is because many of them are known by the controller to be zero at time $t$ since $x_{\tau}$ is assumed to be known, and for each $\tau \in \{t,\ldots,t+T\}$, there is exactly one $i$ such that
\begin{equation*}
i = \argmin_{k \in \{1,\ldots,m\}} \|x_{\tau}-x^k\|_2.
\end{equation*}
The constraint $(s,e,y,z,w) \in \mathcal{F}_t$ enforces the binary restrictions on $z$ and $w$, the system constraints \eqref{powerflow}-\eqref{ybounds}, the constraint
\begin{equation}
\sum_{j=1}^n z^{ij}_{\tau} = 1 \label{zequality} 
\end{equation}
for each $\tau \in \{t,\ldots,t+T\}$, and the constraint
\begin{equation}
z^{ij}_{\tau} = 1 \iff j = \argmin_{k \in \{1,\ldots,n\}} \|y_{\tau}-y^k\|_2 \label{ybinning}
\end{equation}
for each $\tau \in \{t,\ldots,t+T\}$ and $j \in \{1,\ldots,n\}$. Due to \eqref{zequality}, the constraints \eqref{ybounds} and \eqref{ybinning} can be represented together by the linear constraints 
\begin{equation*}
   \sum_{k=1}^{n} z^{ik}_\tau y^k- \Delta \leq y_\tau < \sum_{k=1}^{n} z^{ik}_\tau y^k + \Delta. 
\end{equation*}
Problem \eqref{MIQP} is therefore an MIQP with $(n+1)(T+1)$ binary variables, roughly $3(T+1)$ continuous variables, and roughly $9(T+1)$ linear constraints. Problem \eqref{MIQP} is also recursively feasible \cite{Lofberg2012} as long as the maximum consumer energy demand is less than the maximum allowable grid energy demand, \emph{i.e.}, $X^{\max} \leq Y^{\max}$, and if generation curtailment is allowed. This stems from the fact that the grid can make up for any energy shortfall, local generation in excess of consumption and storage is curtailable, and mutual information is penalised in the objective function and not formulated as a constraint. In this work, generation curtailment is not considered as local generation is assumed to be deterministic and zero; but this can easily be included.

% MDPC
Since this proposed MPC scheme predicts the effects of the controller's actions on the probability distribution of random variables, we refer to it as Model-Distribution Predictive Control, or MDPC. 

%%%%%%%%%%%%%%%%%%%%%%%%%%%%%%%
%%%%%%%%%%%%%%%%%%%%%%%%%%%%%%%
\section{Numerical Experiments}
\label{imp}

% Implementation and computer
The proposed scheme was implemented in YALMIP \cite{Lofberg2004} and MATLAB R2015a using the IBM CPLEX 12.6.3 solver. Simulations were done using consumer load profiles generated with tools from \cite{Pflugradt2013} over a period of approximately one month (30 days) with hourly resolution in a two-tier time-of-use energy pricing environment. The computer used for the simulations was an Intel Core i7-2600 CPU with 3.40 GHz of clock speed, and 16.0 GB of RAM.

% regularization
In order to solve the MIQP problems \eqref{MIQP} reliably and reduce the number of solution candidates, a convex penalty term $r(y)$ based on the $l_1$-norm was added to the objective function. This is given by
\begin{equation*}
   r(y) = \frac{\mu \sigma T^{-1} \|Qy-\bar{y}\|_1}{\gamma(\|Qx-\bar{x}\|_1 + \|Qg-\bar{g}\|_1)+1},
\end{equation*}
where $\sigma$ and $\gamma$ are positive constants, and $y,~x$ and $g$ are grid load, consumer load and local generation vectors used in problem \eqref{MIQP} for time $t$. The matrix $Q$ extracts components that correspond to times $t,...,t+T-1$, and $\bar{y},~\bar{x}$, and $\bar{g}$ denote values of $Y_\tau,~ X_\tau$, and $G_\tau$ for $\tau = t,...,t+T-1$ predicted during the solution of problem \eqref{MIQP} for time $t-1$. This term is normalised for the prediction horizon length, and designed to encourage the controller to choose a set of actions at time $t$ that is similar to those predicted at time $t-1$ in the absence of forecast errors. The values of $\sigma$ and $\gamma$ may be chosen such that $r(y)$ does not dominate the objective function, and hence are dependent on the characteristics of a particular data set. Note that the value of $\gamma$ is negligible for this work as generation is assumed to be zero and the consumer load is perfectly known, \emph{i.e.}, there are no forecast errors. Without the $r(y)$ term, the MIQP solver would spend an excessive amount of time performing branch-and-bound searches and utilize excessive memory resources trying to solve certain instances of \eqref{MIQP}.

% Parameters
Unless stated otherwise, the parameters shown in Table \ref{Param} were used for the simulations. Local generation was set to zero throughout the entire simulation period in order to clearly observe the effects of the controller actions. A battery was used as the energy storage device, and the parameters chosen were those of a Tesla Powerwall.  

% Parameters
\begin{table}[!h]
\renewcommand{\arraystretch}{1.1}
\centering
\caption{Default Simulation Parameters.}
\label{Param}
\begin{tabular}{|lc||lc|} \hline 
$T$:                & $12$      & Reg. Coefficient, $\sigma$:     & $0.11$\\
$M$:                & $120$     & Battery Capacity:     & $6.4$ kWh\\      
$N$:                & $132$     & Battery Power:        & $3.3$ kW\\
$m$:                & $15$      & Battery Efficiencies, $\alpha$: & $96$ \%\\
$n$:                & $15$      & Energy Price (high):  & $24.6$ Rp/kWh\\
$\varepsilon$:      & $0.1$     & Energy Price (low):   & $13.15$ Rp/kWh\\ \hline 
\end{tabular}
\end{table}
\vspace{-0.5cm}

%%%%%%%%%%%%%%%%%%%%%%%%%%%%%%%%%%%%%%%%%%%%%%%%%%%%%%%%%%%%%%%
\subsection{Visualization of Load Profiles and Control Actions}
\label{visualisation}

% Load profile
\begin{figure*}
\centering
\vspace{-0.3cm}
\includegraphics[width=1.88\columnwidth]{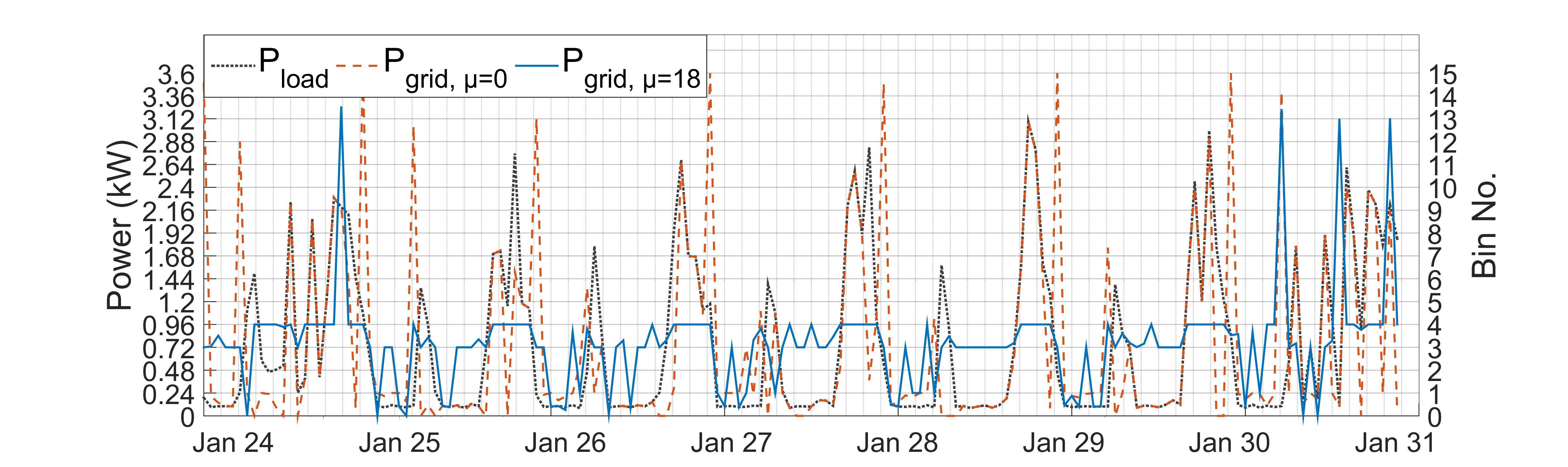}
\vspace{-0.1cm}
\captionsetup{justification=centering}
\caption{Consumer vs grid load.}
\label{LoadvGrid}
\vspace{-0.3cm}
\end{figure*}

% Battery profile
\begin{figure}
\centering
\subfloat[Control actions of the battery]{
\vspace{-0.3cm}
\includegraphics[trim=3cm 0.5cm 3cm 0.5cm, clip=true, width=\columnwidth]{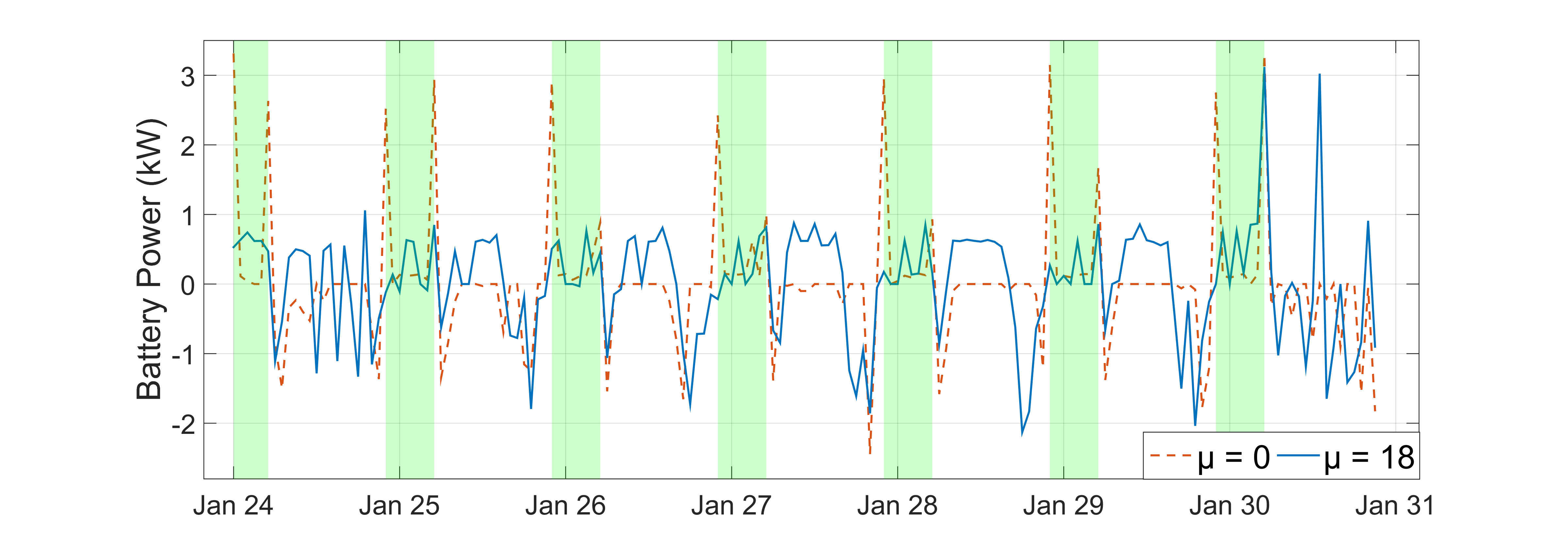}
\vspace{-0.3cm}
\label{battcurve1}
}\\
\subfloat[Battery state of charge]{
\vspace{-0.3cm}
\includegraphics[trim=3cm 0.5cm 3cm 0.5cm, clip=true, width=\columnwidth]{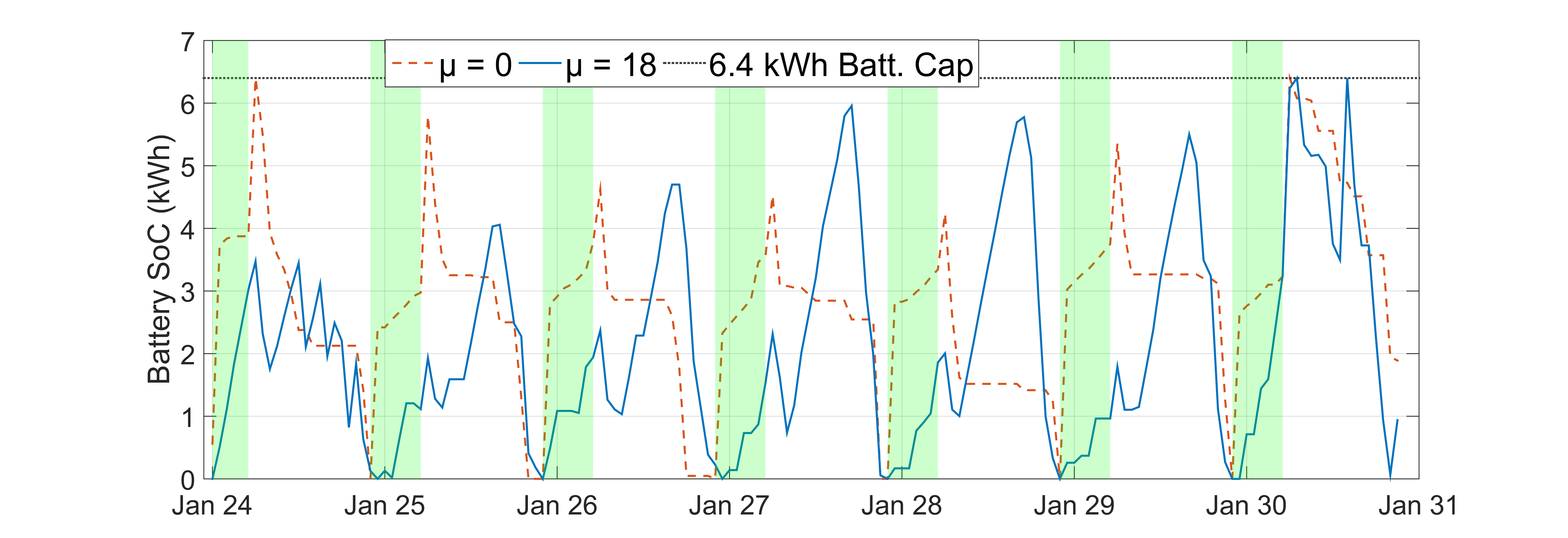}
\label{battcurve2}
}
\vspace{-0.1cm}
\captionsetup{justification=centering}
\caption{Control actions of the battery and its state of charge.}
\end{figure}

\begin{figure}
\centering
\vspace{-0.3cm}
\includegraphics[trim=3.5cm 1cm 3.5cm 1cm, clip=true, width=\columnwidth]{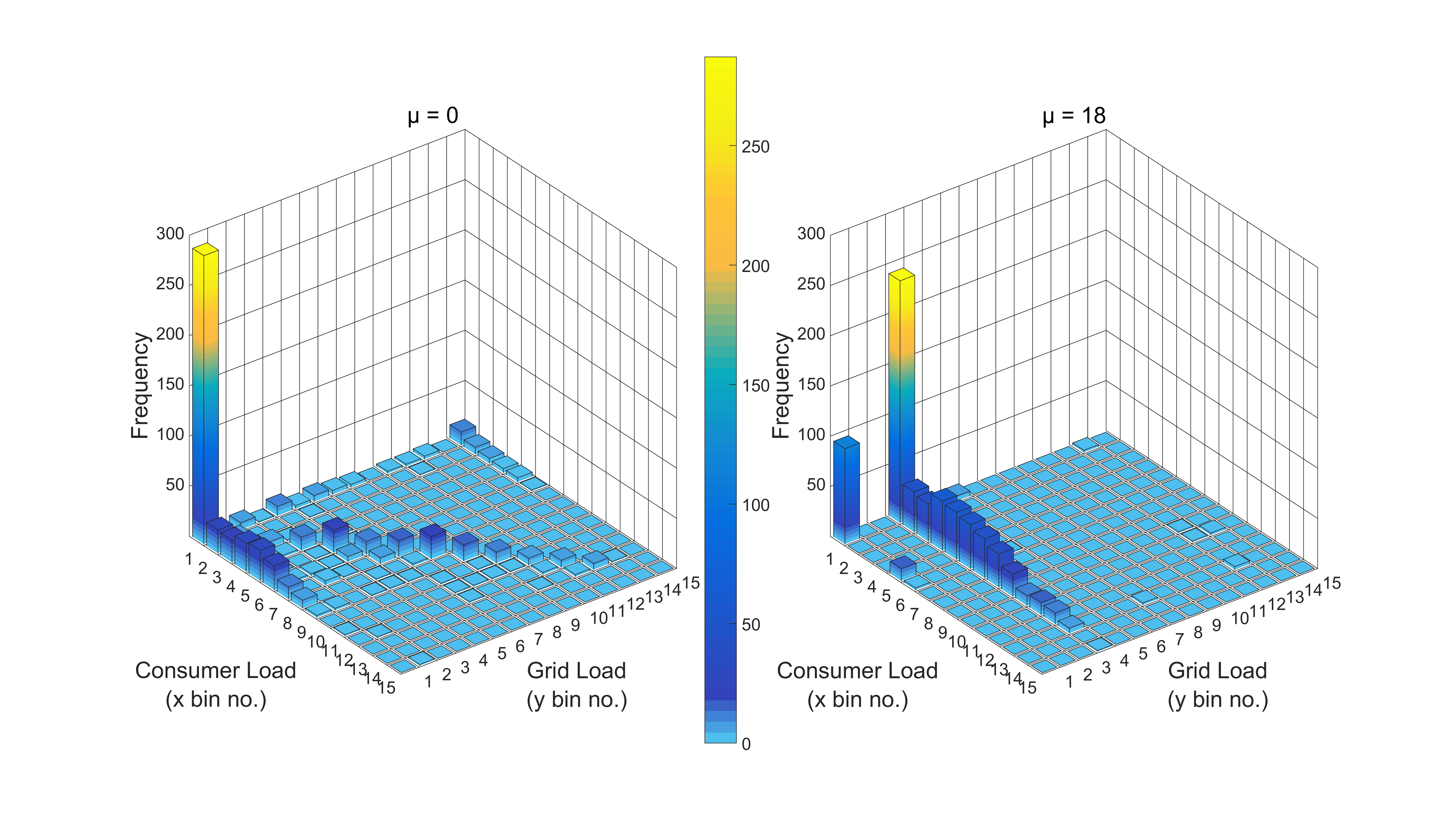}
\vspace{-0.6cm}
\captionsetup{justification=centering}
\caption{Distribution of consumer and grid loads.}
\label{xyplot}
\vspace{-0.1cm}
\end{figure}

% Correlation observations
Fig. \ref{LoadvGrid} illustrates the actual consumer load and the load seen by the grid over seven days for two different prices of privacy loss, $\mu = 0$ and $\mu = 18$. For $\mu=0$, the controller optimizes for energy costs only, and thus charges the battery during low-price periods and compensates consumer load during the high-price periods. This can be seen in Fig. \ref{battcurve1} and \ref{battcurve2}, which show the charge and discharge of the battery, and the battery state of charge over a period of seven days. Periods highlighted in green indicate low-price periods. With the chosen horizon length of $12$, the controller is unable to fully utilize the battery's capacity as the consumer load ``seen'' by it is generally less than the battery's capacity. With a price of privacy loss of $\mu=18$, the controller now charges over high-price periods as well, as seen in Fig. \ref{battcurve1}, in order to mask low consumer load periods. This results in a grid load curve that shows a stepping behavior similar to that obtained in \cite{Yang2012}, with periods of load leveling similar to those obtained in \cite{McLaughlin2011}. Note that the correlation and thus also the mutual information of consumer and grid loads is greatly reduced for $\mu = 18$. This can be seen more clearly in Fig. \ref{xyplot}, which illustrates the estimated distribution of $(\tilde{X}_t,\tilde{Y}_t)$. For $\mu=18$ (right plot), the grid load concentrates in the two levels despite consumer load levels being spread out. For $\mu=0$, while correlation between $\tilde{X}_t$ and $\tilde{Y}_t$ is reduced by the battery charging during low-price hours to compensate consumption during high-price hours, it is still possible for an adversary to reconstruct the consumer load based on observations of the grid load with knowledge of energy prices and the battery system specifications, as discussed in \cite{Yang2012}.

% Joint distribution

%%%%%%%%%%%%%%%%%%%%%%%%%%%%%%%%%%%%%%%%%%%
\subsection{Performance of the MDPC Scheme}
\label{performance}

% Cumulative I 
The proposed scheme was evaluated using $\bar{I}(\tilX_t;\tilY_t)$, equivalently \eqref{nonlinear}, with a static window of length $N=718$ that encompasses the entire simulation period of approximately 30 days. This ``cumulative" approximation of mutual information is denoted by $I_c$. For reference, a system without a battery using the same simulation setup had a total energy consumption of $534$ kWh, energy costs of $125$ CHF, and privacy loss of $I_c = 2.58$ bits. 

% I and cost vs mu
\begin{figure}
\centering
\subfloat[Mutual information]{
\includegraphics[trim=3cm 0cm 3cm 0.5cm, clip=true, width=\columnwidth]{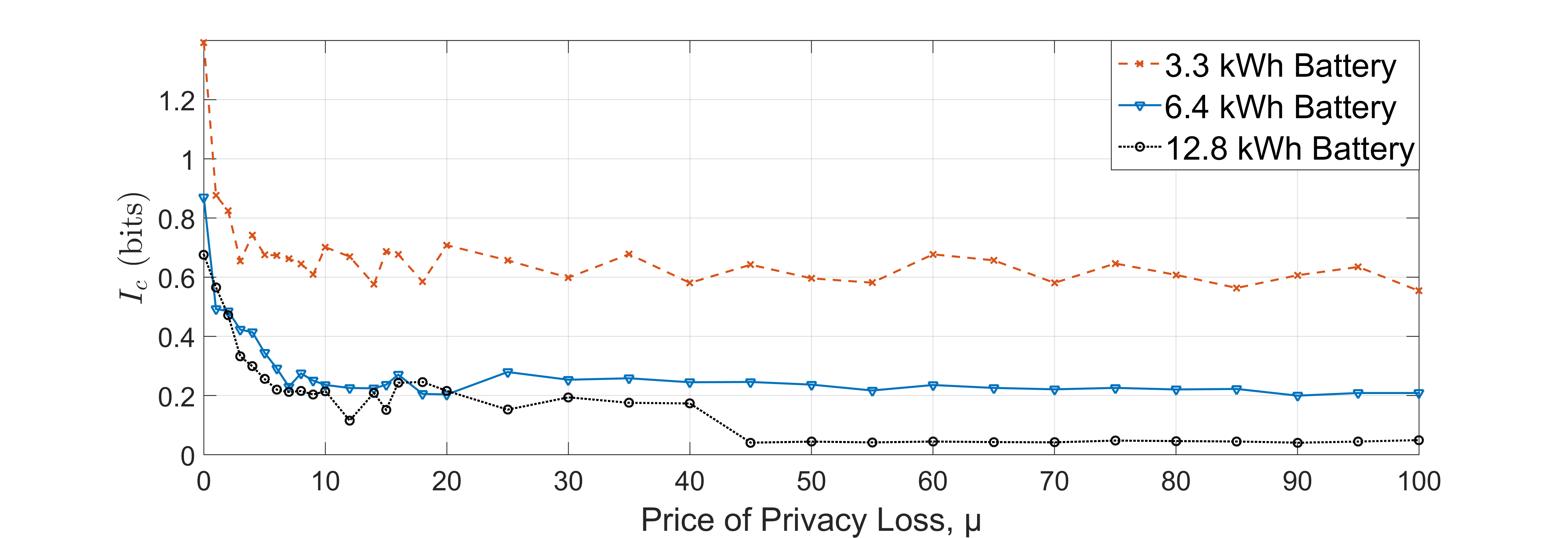}
\label{iccurve}
}\\ \vspace{-0.3cm}
\subfloat[Total energy cost]{
\includegraphics[trim=3cm 0cm 3cm 0.5cm, clip=true, width=\columnwidth]{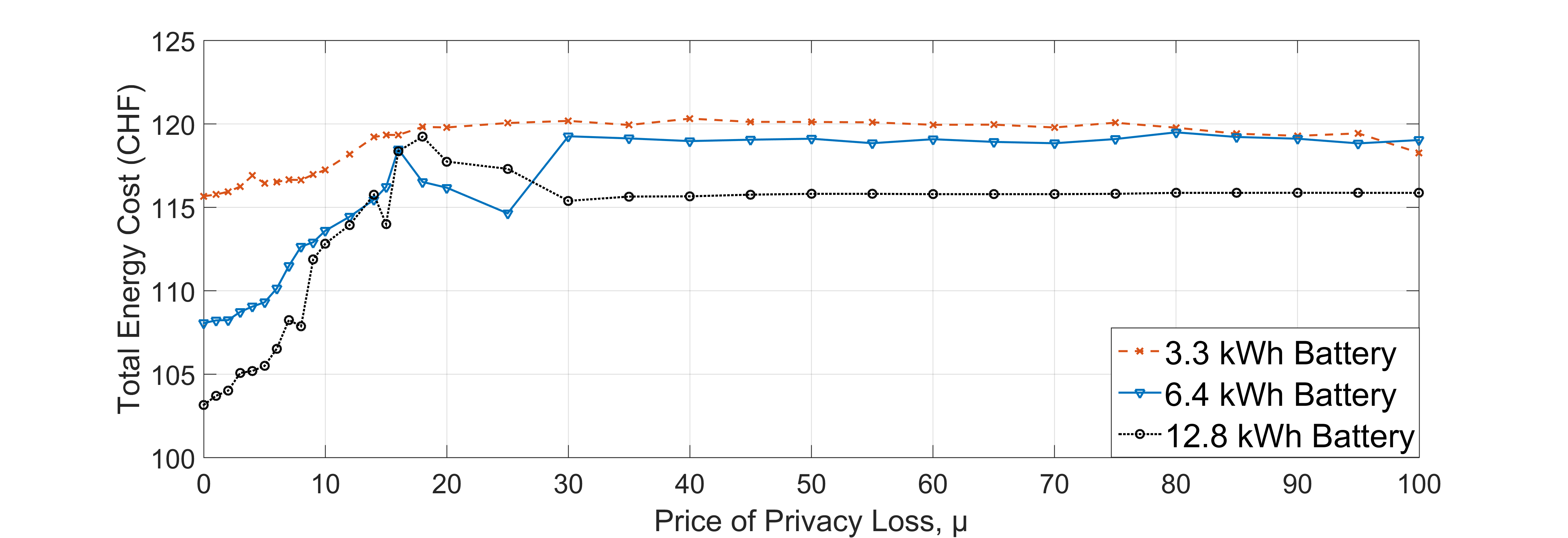}
\label{costcurve}
} \vspace{-0.1cm}
\captionsetup{justification=centering}
\caption{Comparison of different battery capacities.}
\label{iccostcurves}
\end{figure}

% Fluctuations
\begin{figure}
\centering
\includegraphics[trim=2.5cm 0cm 3cm 0.5cm, clip=true, width=\columnwidth]{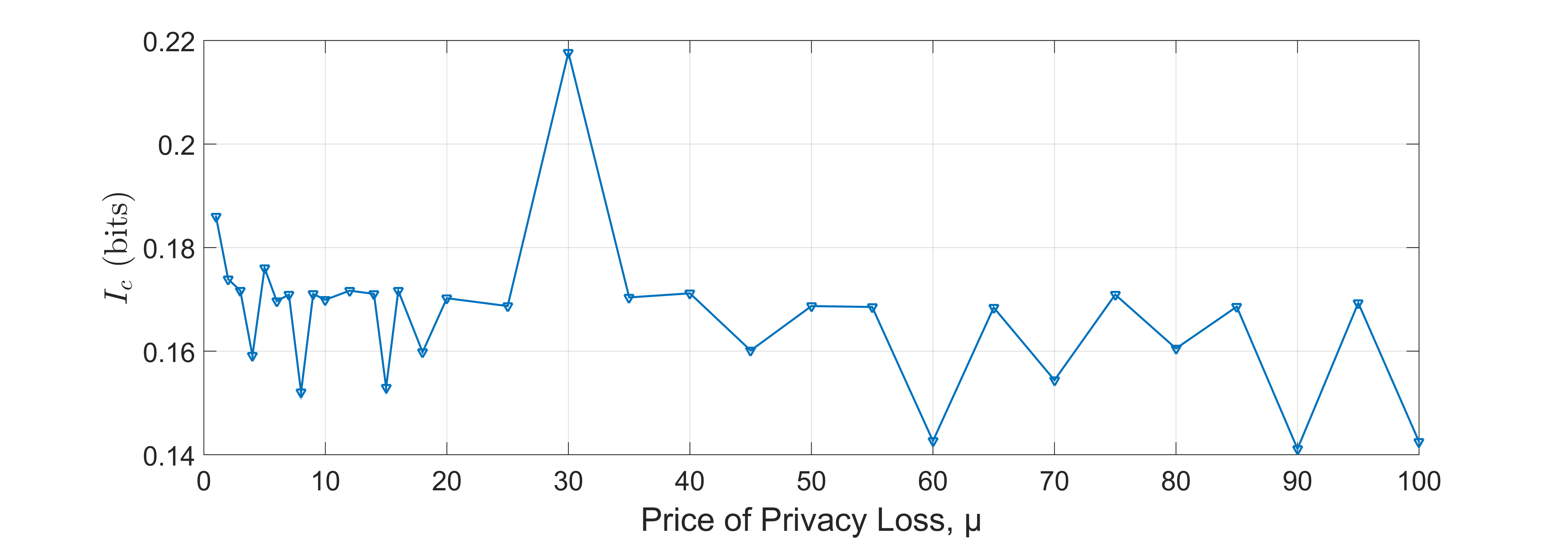}
\vspace{-0.3cm}
\captionsetup{justification=centering}
\caption{Mutual information when energy cost is ignored.}                
\label{randdevcurves}
\end{figure}

% I and cost vs mu analysis
In Fig. \ref{iccostcurves}, the total energy costs and mutual information $I_c$ for different battery capacities are shown as a function of the price $\mu$ of privacy loss. As can be seen from the figure, energy costs increase with the drop in mutual information, as the battery charges during high-price periods to mask low consumer loads, as previously shown in Fig. \ref{battcurve1}. While it would be possible for consumer behavior to be inferred with $\mu=0$, the mutual information value with a $3.3$ kWh battery for energy cost optimization only is already approximately half that of the no-battery case.   

Similar to the findings in \cite{Yang2012}, the size of the battery relative to the total consumer load over a high-low consumption cycle determines the level of privacy protection that can be achieved. Privacy loss is minimal with a battery capacity that sufficiently covers the consumer load during a typical day. In the case of the simulated consumer load, this battery capacity is approximately $6.4$ kWh, with a $12.8$ kWh battery only achieving slightly better performance in terms of privacy loss reduction for moderate values of $\mu$. However, energy cost reduction improves with a larger battery, since it also allows covering the consumption during periods of high energy price and low load. 

% Fluctuations
As seen in Fig. \ref{iccurve}, minimum privacy loss is achieved with a price of privacy loss of about $20$ Rp/bit for a $6.4$ kWh battery. Beyond this price of privacy loss, the system limitations (or the limitations of the measure $I_c$) impede the controller from doing better, and the mutual information is seen to fluctuate around the minimum value. These fluctuations are caused by the inherent randomness that is present in the scheme, as well as due to the finiteness of the simulation period. This is seen Fig. \ref{randdevcurves}, which shows how $I_c$ varies as the price of privacy loss increases even when the energy cost is ignored. It can be seen that the ``width'' of these fluctuations is roughly $0.1$ bits. They arise in part from the MIQP solver finding multiple solution candidates and choosing different ones, and hence affecting the system trajectory, purely due to the scaling of the objective function, which is completely controlled by $\mu$ when the energy cost is ignored. 

\begin{figure}
\centering
\subfloat[Discretization levels]{
\includegraphics[trim=3cm 0cm 3cm 0.5cm, clip=true, width=\columnwidth]{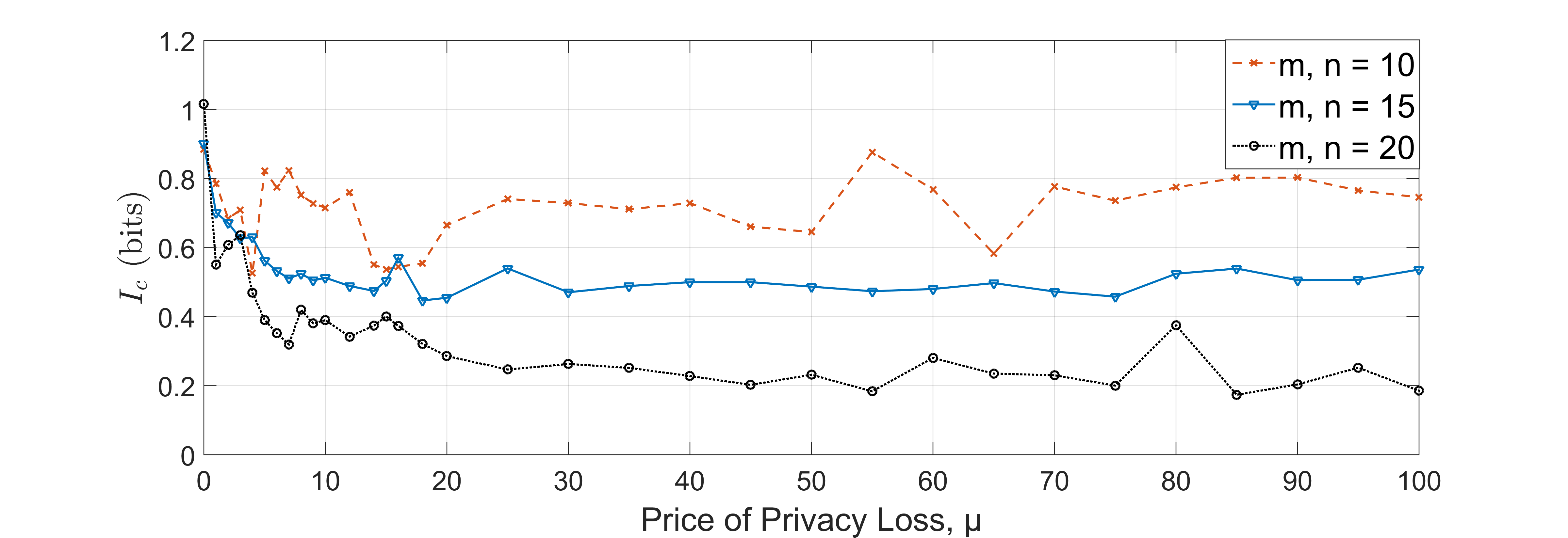}
\label{bincomp}
}\\
\subfloat[Battery power ratings]{
\includegraphics[trim=3cm 0cm 3cm 0.5cm, clip=true, width=\columnwidth]{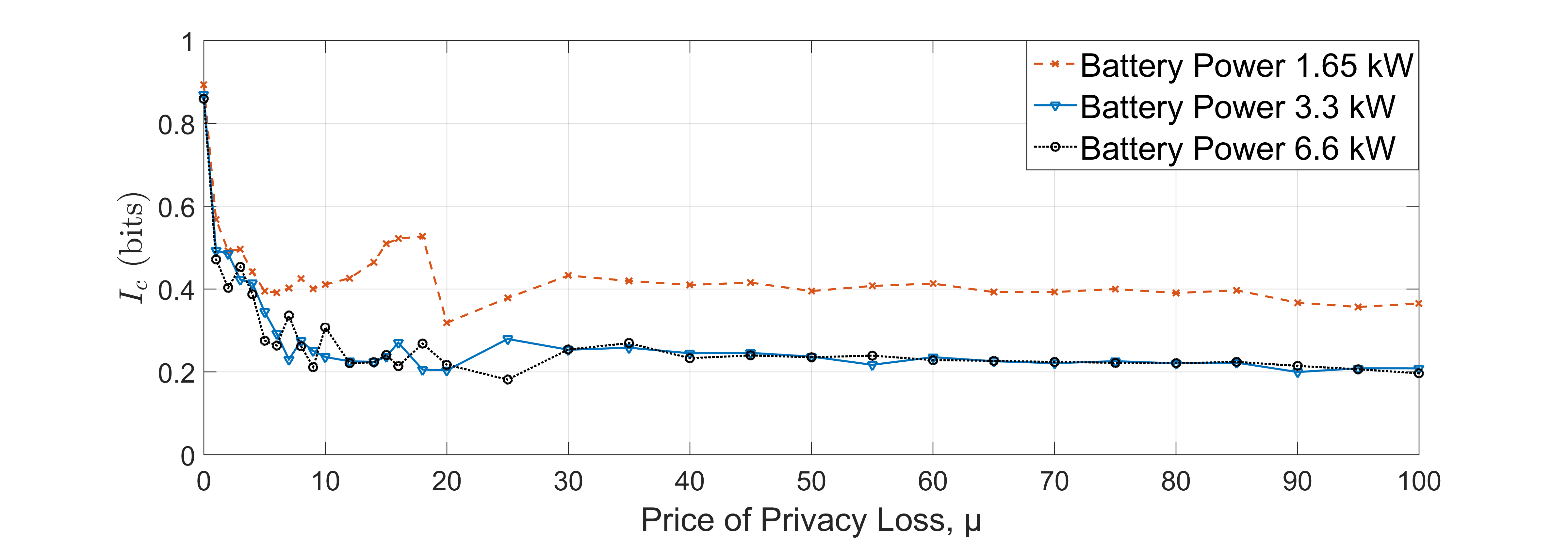}
\label{powercomp}
}
\captionsetup{justification=centering}
\caption{Effects of discretization levels and battery power ratings on mutual information.}
\label{binpowercomp}
\vspace{-0.6cm}
\end{figure}

The effect of the number of quantization levels of $\tilX_t$ and $\tilY_t$, which are $m$ and $n$, respectively, and of the charge-discharge rate limit of the battery on the performance of the proposed controller were also studied. Fig. \ref{binpowercomp} shows $I_c$ as a function of $\mu$ for different number of quantization levels and battery power ratings. For the evaluation of the quantization levels, $(m,n)=(20,20)$ was used. Increasing the number of quantization levels for the same maximum load allows the controller to potentially achieve lower levels of privacy loss because it is now able to take advantage of the higher resolution to fine-tune its control actions. However, this also generally increases the fluctuation amplitude or variability of $I_c$ due to the MIQP solver finding a larger set of candidate solutions. Nonetheless, smaller quantization levels could also result in large fluctuations when evaluated at higher resolutions due to the coarse level of controller actions, such as in the case of $(m,n) = (10,10)$. The battery power ratings on the other hand impose a limit on the controller's ability to mask changes in consumer load. For the considered simulation setup with a maximum consumer load of $3.6$ kW and a battery energy capacity of 6.4 kWh, a battery rating of $3.3$ kW is almost sufficient. This can be seen in Fig. \ref{powercomp}, where a battery with a $6.6$ kW power rating marginally improves performance. A battery with a $1.65$ kW rating is sufficient for lower prices of privacy loss, as energy cost reduction is the main driver of control actions in such cases. However, at higher prices of privacy loss the controller is unable to fully mask the consumer load peaks that occur mainly during high-energy-price periods, as shown in Fig. \ref{battcurve1}.

% Bin analysis

% Horizons
\begin{table*}
\renewcommand{\arraystretch}{1.1}
\caption{Comparison Between Different Horizon Lengths.}
\label{Computational}
\centering
\begin{tabular}{|c|ccc|ccc|ccc|}
\hline 
& \multicolumn{3}{|c|}{$\mu = 0$}   &\multicolumn{3}{c|}{$\mu = 15$}    &\multicolumn{3}{c|}{$\mu = 45$} \\ 
\hline 
                            & $T = 12$  & $T = 18$  & $T = 24$  & $T = 12$  & $T = 18$  & $T= 24$   & $T = 12$  & $T = 18$  & $T= 24$ \\ 
\hline 
Max Solver Time (s)         & $0.072$   & $0.085$   & $0.112$   & $1.14$    & $11.0$    & $152$     & $1.39$    & $11.4$    & $211$ \\ 
\hline 
Min Solver Time (s)         & $0.024$   & $0.027$   & $0.036$   & $0.053$   & $0.109$   & $1.59$    & $0.053$   & $0.092$   & $1.28$ \\
\hline 
Mean Solver Time (s)        & $0.031$   & $0.039$   & $0.048$   & $0.161$   & $1.03$    & $9.26$    & $0.152$   & $0.961$   & $10.3$ \\
\hline 
Median Solver Time (s)      & $0.030$   & $0.036$   & $0.045$   & $0.126$   & $0.712$   & $5.64$    & $0.117$   & $0.667$   & $5.60$ \\ 

\hline 
\end{tabular} 
\end{table*}

%Horizon Costs and Ic
\begin{figure}
\centering
\subfloat[Mutual information]{
\includegraphics[trim=3cm 0cm 3cm 0.5cm, clip=true, width=\columnwidth]{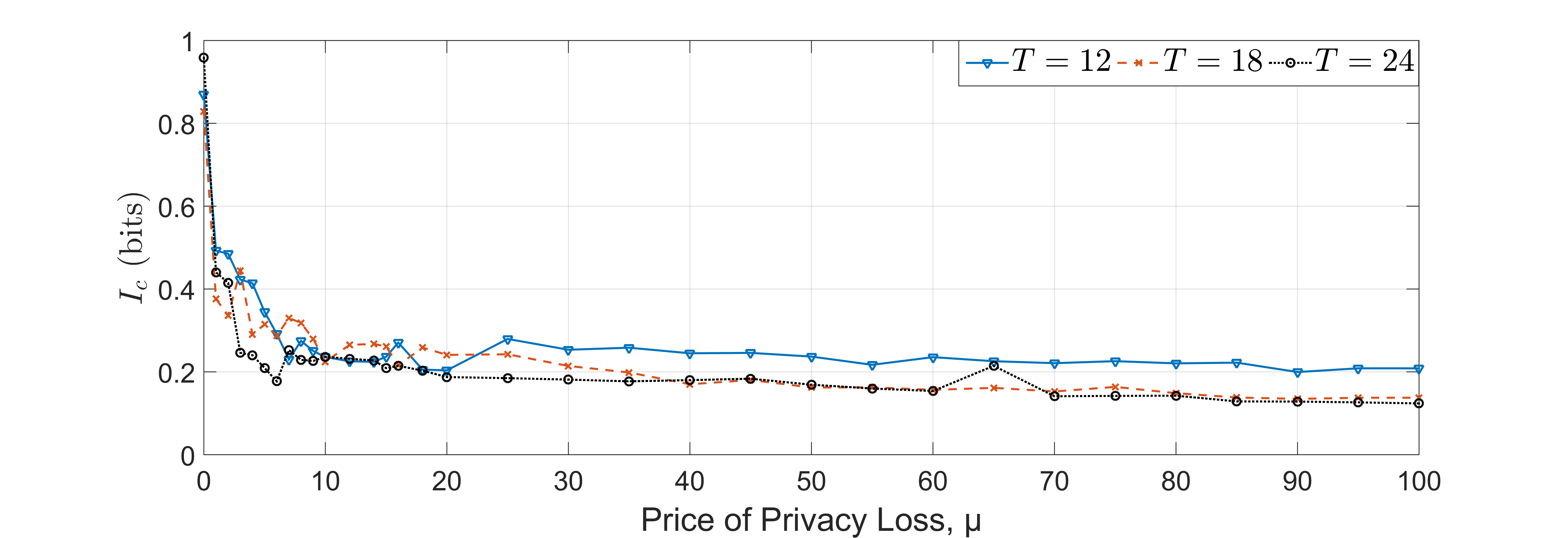}
\label{horizonic}
}\\
\subfloat[Total energy costs]{
\includegraphics[trim=3cm 0cm 3cm 0.5cm, clip=true, width=\columnwidth]{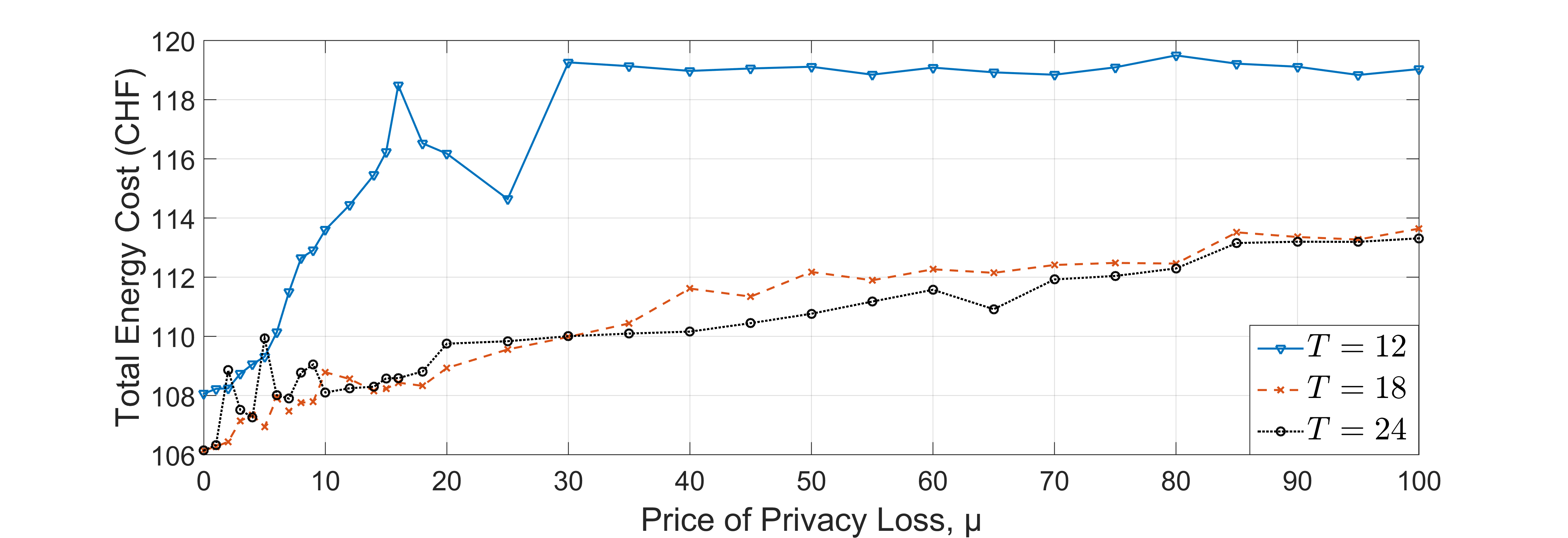}
\label{horizoncosts}
}
\captionsetup{justification=centering}
\caption{Comparison of different prediction horizon lengths}
\label{horizoncompare}
\vspace{-0.3cm}
\end{figure}

% Horizons
Additionally, as discussed in Section \ref{visualisation}, a horizon length of $T=12$ is insufficient for the controller to predict the load and fully charge the battery to minimize energy costs and mask consumer loads during the later stage of high energy price periods. Hence, the controller was simulated with different horizon lengths to study its performance and computational requirements. Table \ref{Computational} and Fig. \ref{horizoncompare} summarize the results obtained. As can be seen from Fig. \ref{horizoncompare}, prediction length mainly affects the energy cost because it allows the controller to anticipate consumer load that occur farther in the future. Increasing $T$ to $18$ reduces energy costs, and generally also reduces privacy-loss, due to the controller being able to fully utilize the capacity of the $6.4$ kWh battery. However, there are exceptions to this due to the randomness in the scheme, which is exacerbated by longer horizons since they enlarge the solution space and increase the number of candidate MIQP solutions. It is important to note that computational tractability is still maintained with $T =24$ despite having a median solver time of 57 times that of $T=12$. However, as seen in Table \ref{Computational}, the maximum solver time increases exponentially with the the prediction horizon. Further increasing the horizon length to $T=30$ makes it computationally impractical for a realistic controller as it requires more than one and a half hours using the current implementation to solve a single MIQP problem, which needs to be completed in under an hour to be practical. Note that for the energy pricing structure and consumer load profile in the simulation setting, a horizon length of $18$ appears to be sufficient. However, this should not be generalized to other simulation settings and load profiles. In general, using the scheme proposed, a controller that is able to predict longer time periods while maintaining tractability should perform better.

%%%%%%%%%%%%%%%%%%%%%%%%%%%%%%%%%%%%
%%%%%%%%%%%%%%%%%%%%%%%%%%%%%%%%%%%%
\subsection{Time Evolution of Mutual Information} 
\begin{figure}
\centering
\includegraphics[trim=2.5cm 0cm 3cm 0.5cm, clip=true, width=\columnwidth]{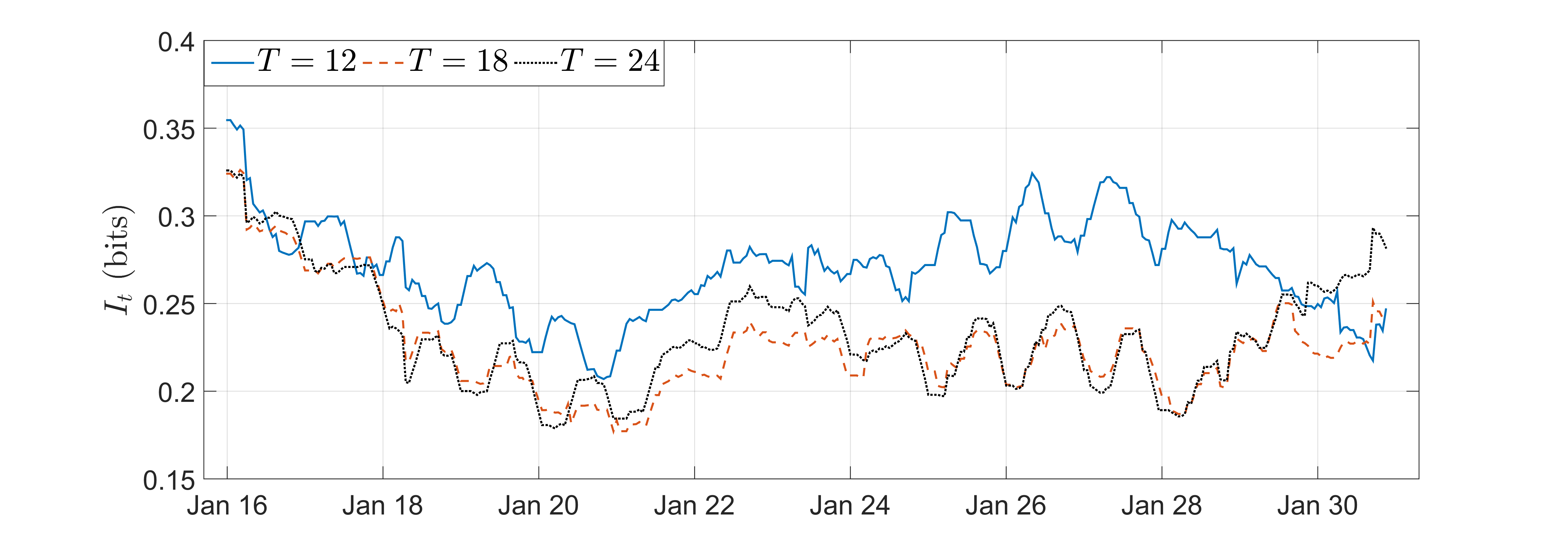}
\captionsetup{justification=centering}
\vspace{-0.3cm}
\caption{Time series of mutual information with different prediction horizon lengths}
\label{ixytimeseries}
\end{figure}

\begin{figure}
\centering
\includegraphics[trim=2.5cm 0cm 3cm 0.5cm, clip=true, width=\columnwidth]{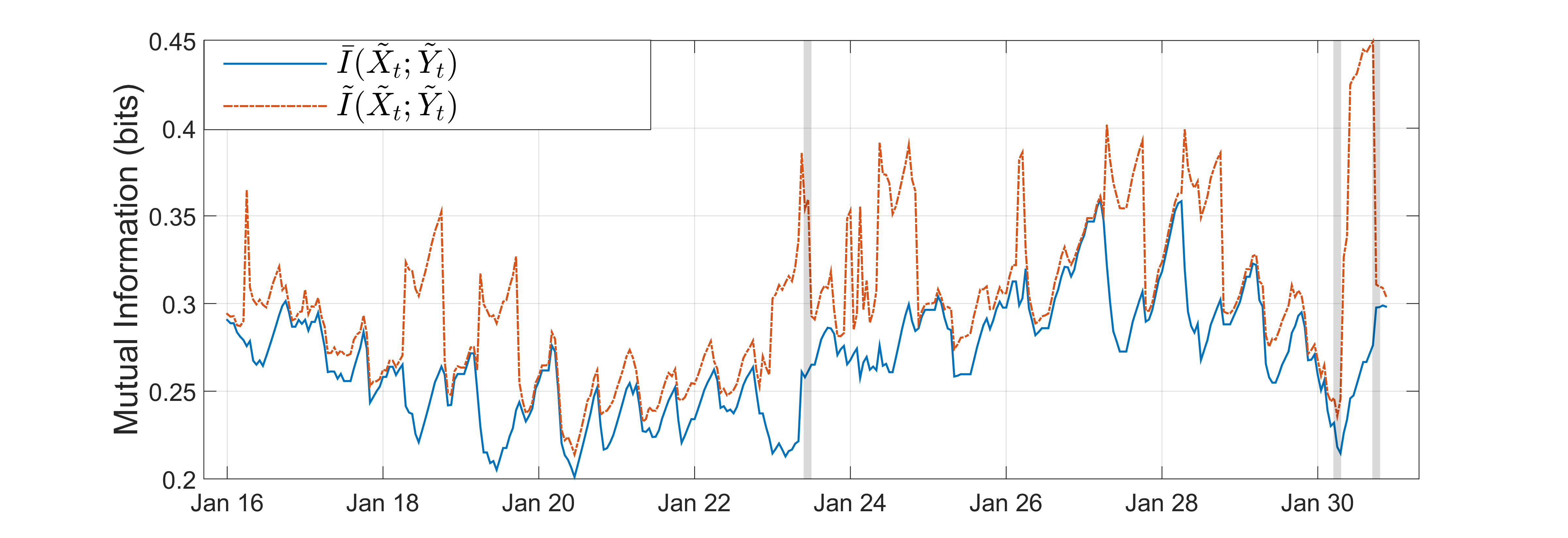}
\captionsetup{justification=centering}
\vspace{-0.3cm}
\caption{Effects of logarithmic linearization}
\label{ixytstep}
\end{figure}

In order to study the change in mutual information with time, we evaluated the controller's performance with $\mu=45$ for the three different horizon lengths discussed in Section \ref{performance} using $\bar{I}(\tilX_t;\tilY_t)$, \emph{i.e.}, \eqref{nonlinear}, with a moving window of length $N=132$. This ``time-dependent" approximation of mutual information is denoted by $I_t$. For each time $t$, the window considered the observation history from time $t-N+1$ to time $t$, as shown in Fig. \ref{SamplingTime}, hence it was based on the actual implemented controller actions. The graphs obtained of $I_t$ are shown in Fig. \ref{ixytimeseries}. As seen from the plot, the results are consistent with the findings in Section \ref{performance}, where longer horizon lengths also lead to generally lower mutual information over time. Note that mutual information does not change monotonically over time due to changing consumer load, the inherent randomness in the scheme, and the logarithm linearizations done by the controller, which make it overestimate the mutual information to varying degrees depending on the statistics of $\tilX_t$ and $\tilY_t$. The latter can be seen in Fig. \ref{ixytstep}, which now shows $\bar{I}(\tilX_t;\tilY_t)$ with the time window $\{t-M+1,\ldots,t+T\}$ for $\mu =20$ and $T=12$, and the approximation with linearized logarithms used by the controller, \eqref{linlog}, which is denoted by $\tilI(\tilX_t;\tilY_t)$. As seen on the plot, at certain times, the discrepancy introduced by the linearization leads to the controller taking actions that are expected to reduce mutual information, but in fact increase it. For example, this can be seen around the time corresponding to $11$ AM on the $23$rd of January and at $5$ PM on the $30$th of January. Around $6$ AM on the $30$th of January, on the other hand, this is reversed: mutual information is reduced when the controller predicted actions that supposedly increased mutual information. These periods are highlighted in grey on Fig. \ref{ixytstep}.

% I_t curve

% I Step curve

\subsection{Comparison of MDPC and Load-Leveling}
\begin{figure}
\centering
\subfloat[3.2 kWh battery]{
\includegraphics[trim=3cm 0cm 3cm 0.5cm, clip=true, width=\columnwidth]{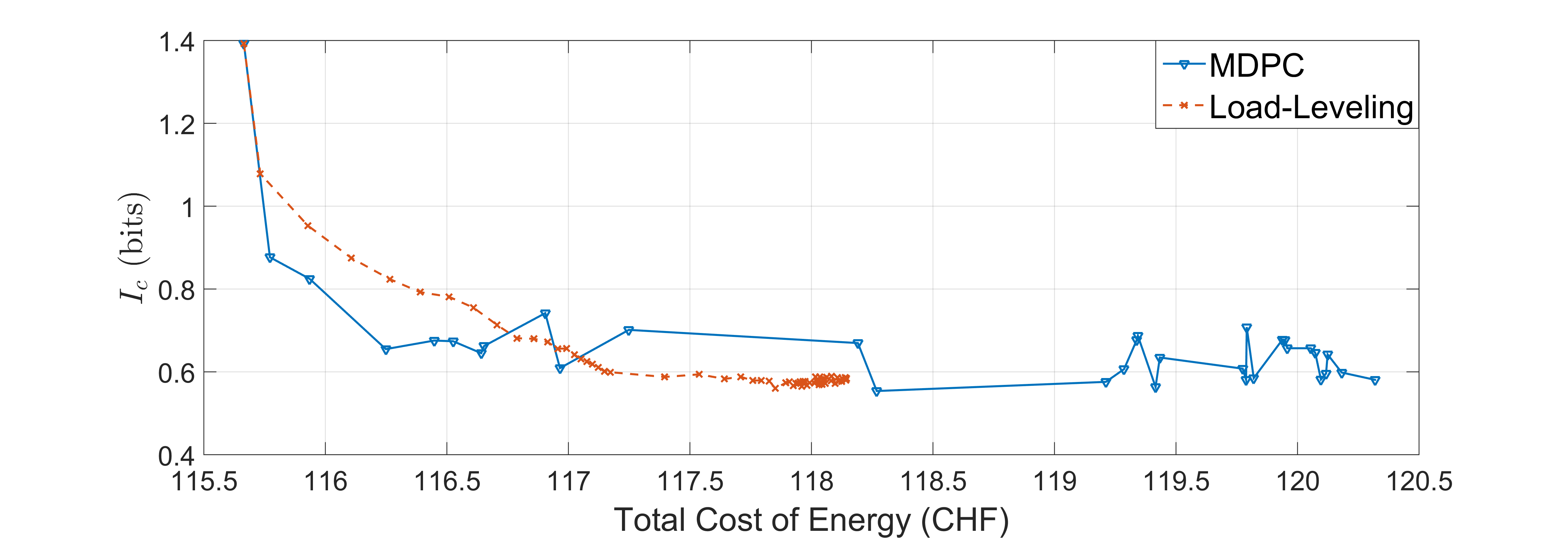}
\label{nillcomp32}
}\\
\subfloat[6.4 kWh battery]{
\includegraphics[trim=3cm 0cm 3cm 0.5cm, clip=true, width=\columnwidth]{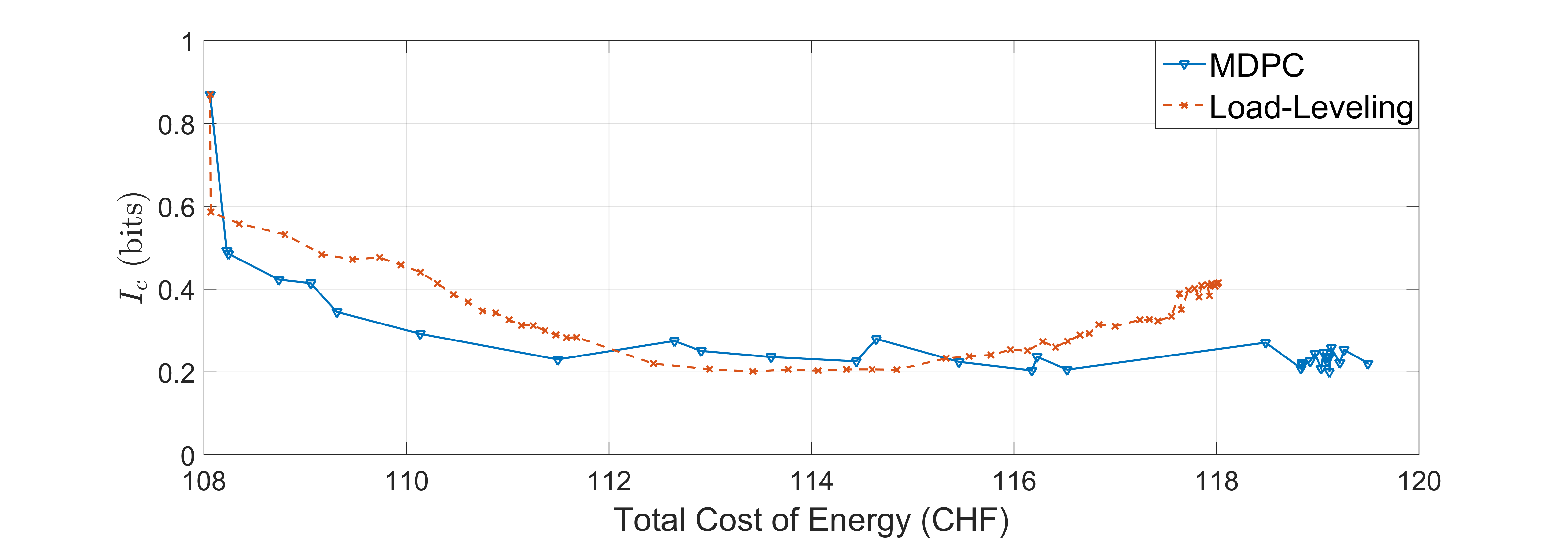}
\label{nillcomp64}
}\\
\subfloat[12.8 kWh battery]{
\includegraphics[trim=3cm 0cm 3cm 0.5cm, clip=true, width=\columnwidth]{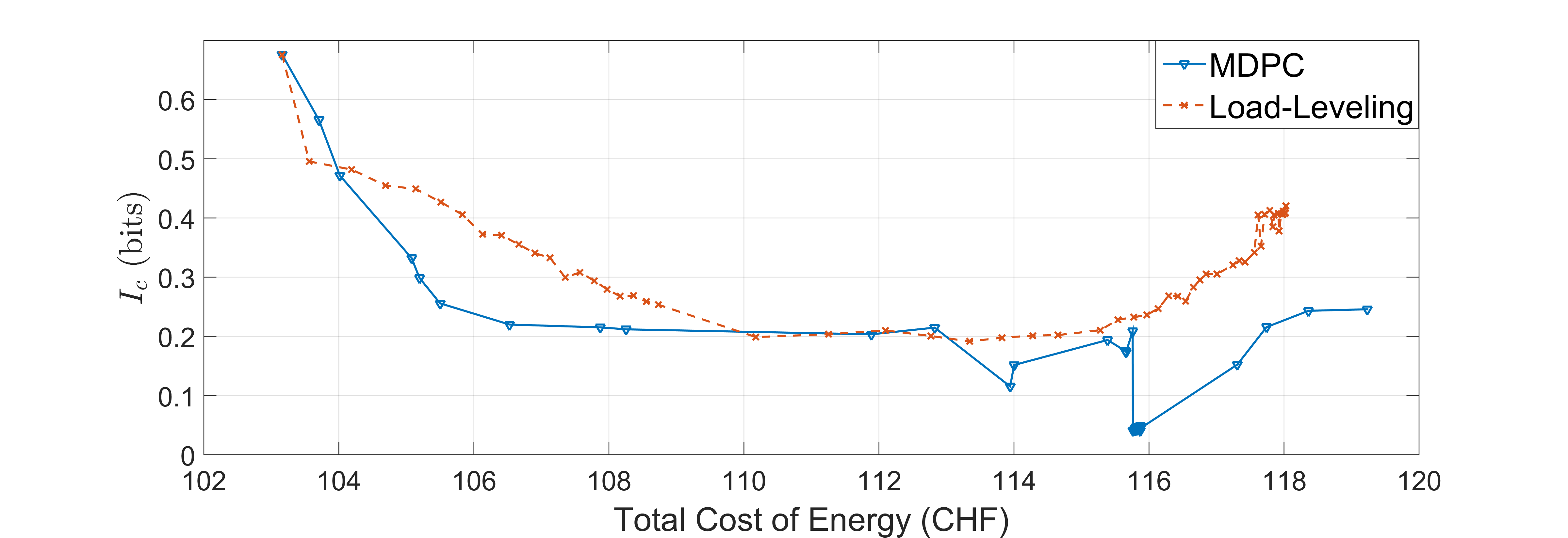}
\label{nillcomp128}
}
\captionsetup{justification=centering}
\caption{Trade-off between $I_c$ and energy cost of MDPC and Load-Leveling with different battery capacities}
\label{nillcomp}
\end{figure}

In the interest of comparing the performance of the proposed MDPC scheme against other BLH approaches, we implemented a load-leveling scheme similar to the heuristic-driven approach in \cite{McLaughlin2011}, but in an MPC setting with a dual objective to also minimise energy cost. More specifically, an MPC problem for this scheme consists of
\begin{alignat*}{2}
& \underset{s,e,y,w}{\mbox{minimize}} \quad && \frac{1}{T+1} \sum_{\tau=t}^{t+T} c_{\tau} y_{\tau} + \frac{\mu}{T+1} \sum_{\tau=t}^{t+T} (y_{\tau}-y_{\tau-1})^2 \\
& \mbox{subject to} \quad && (s,e,y,w) \in \bar{\mathcal{F}}_t, 
\end{alignat*}
where $(s,e,y,w) \in \bar{\mathcal{F}}_t$ enforces the system constraints \eqref{powerflow}, \eqref{powerlimits}, \eqref{dynamics} and \eqref{batterycap}, and the binary restrictions on $w$. Fig. \ref{nillcomp} illustrates the trade-off between mutual information and energy cost for both the MDPC and Load-Leveling (LL) schemes for different battery capacities. As can be seen in Fig. \ref{nillcomp}, both schemes have increased energy costs with the reduction in mutual information, which is expected. When balancing between mutual information and energy cost, the MDPC scheme out-performs load-leveling, generally achieving lower mutual information for a similar cost of energy. 

\begin{figure}
\subfloat[MDPC with $\mu=20$]{
\includegraphics[trim=1.8cm 0cm 1.8cm 0.5cm, clip=true, width=\columnwidth]{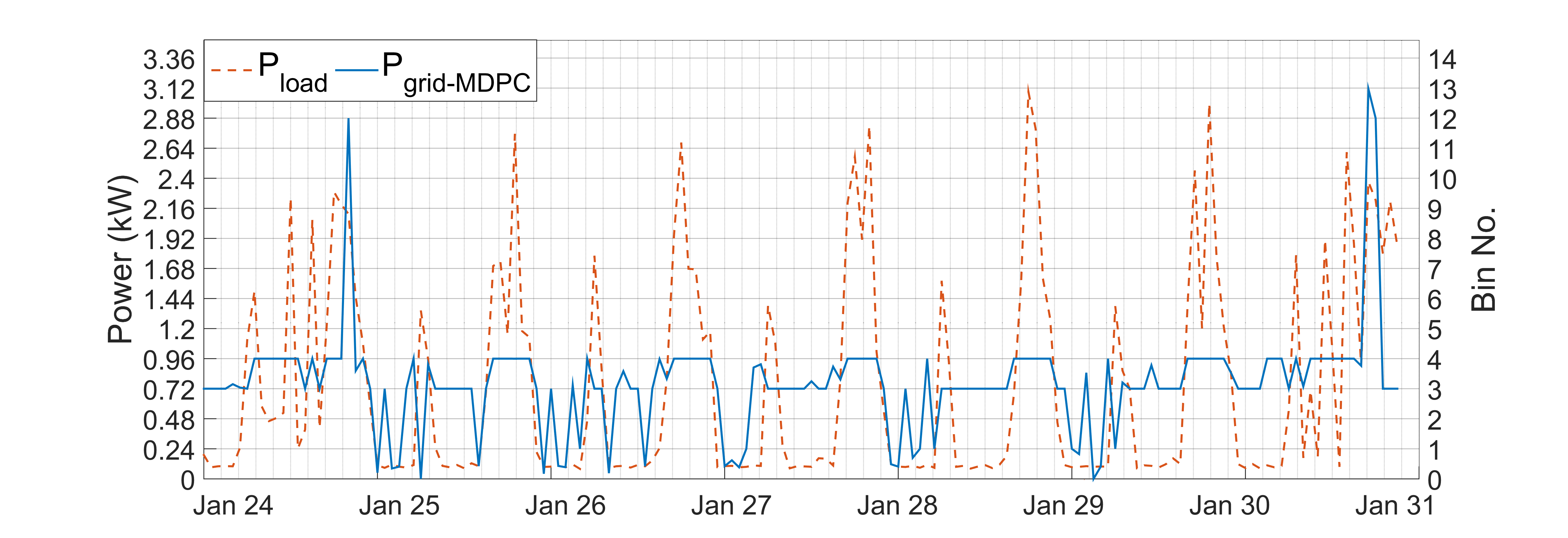}
\label{loadvgridmdpc}
}\\
\subfloat[Load-Leveling with $\mu=30$ and $1080$]{
\includegraphics[trim=1.8cm 0cm 1.8cm 0.5cm, clip=true, width=\columnwidth]{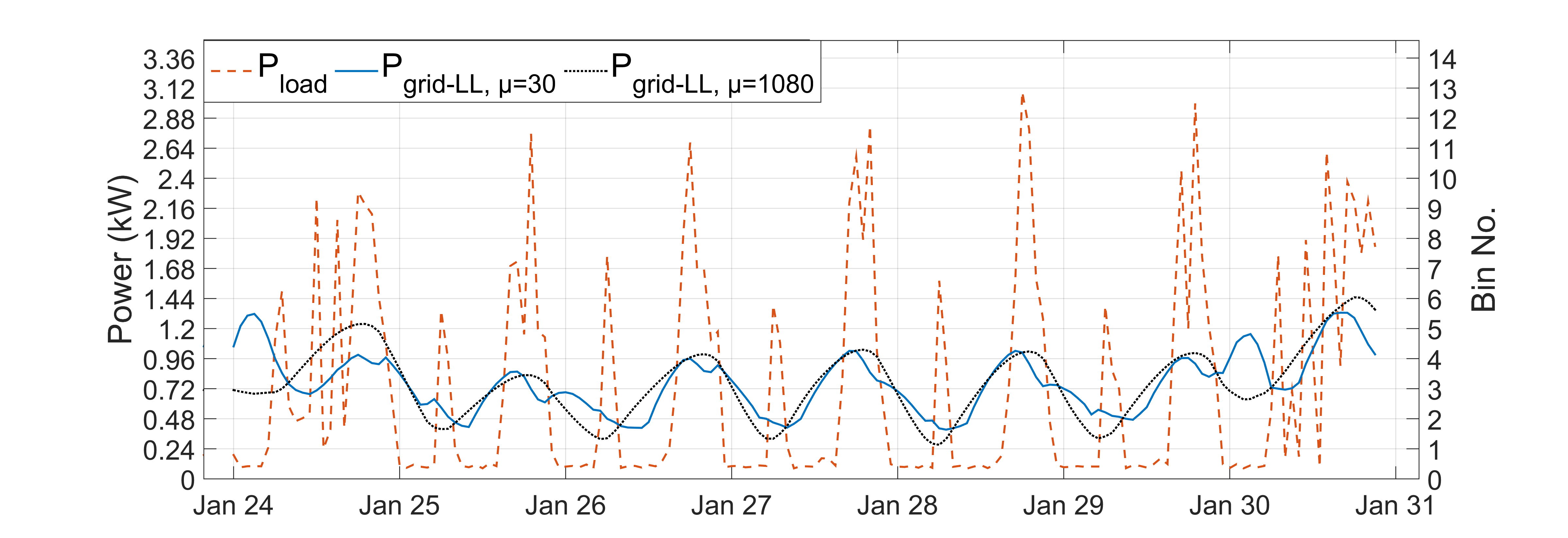}
\label{loadvgridnill}
}
\captionsetup{justification=centering}
\caption{Consumer vs grid load curves of MDPC and Load-Leveling}
\label{mdpcnillloadvgrid}
\end{figure}

While both schemes are able to achieve similar levels of minimum privacy loss according to the proxy $I_c$ for the different battery capacities, the resultant grid-load curves are quite different, as seen in Fig. \ref{mdpcnillloadvgrid}. The grid load curve for the MDPC scheme shown in Fig. \ref{loadvgridmdpc} no longer resembles the diurnal pattern of the actual consumer load, whereas this pattern can be clearly discerned from the load-leveling grid load curves shown in Fig. \ref{loadvgridnill}, which suggest that they should leak more information. These results strongly suggest that the mutual information proxy used by the MDPC scheme and for the evaluations is unable to capture these important properties, which is expected since they are related to time correlation. It is interesting (and perhaps surprising) that despite the proxy not being able to fully capture time correlation, the MDPC scheme is able mask the diurnal pattern in consumer load profiles. 

% Trade off curves

% Load vs Grid Curves

%%%%%%%%%%%%%%%%%%%%
%%%%%%%%%%%%%%%%%%%%
\section{Conclusion}
\label{Conclusion}

% Summary
In this paper, we proposed an MPC-based controller that balances energy costs and privacy protection based on mutual information. This is done by predicting the effects of the controller's actions on the statistics of the consumer load and that seen by the grid using counting, and solving MIQP problems of manageable size whenever new meter readings are available. While not computationally scalable with prediction horizon and discretization levels of load seen by the grid, the results obtained showed that the controller is able to reduce information leakage and hence protect consumer privacy. Extensive experiments were carrier out in this work to thoroughly characterize the performance of the proposed controller. In particular, the effects of battery size, power ratings, prediction horizon lengths,  and discretization were studied. Comparing the scheme against load-leveling also indicated an improvement in balancing between energy cost and consumer privacy protection. An important result of this work is that it shows that practical schemes that specifically target the minimization of a rigorous measure of privacy loss, such as the one proposed here, are possible.

% Future research 
Future work will focus on enlarging the reach of the prediction horizon, accounting for time correlation of the load consumptions, including more sophisticated probability estimation techniques, comparing the approach to other BLH schemes, and exploring MIQP relaxation methods to make the scheme more computationally scalable.

\bibliographystyle{IEEEtran}
\bibliography{./IEEETrans}

% Generated by IEEEtran.bst, version: 1.14 (2015/08/26)
\begin{thebibliography}{10}
\providecommand{\url}[1]{#1}
\csname url@samestyle\endcsname
\providecommand{\newblock}{\relax}
\providecommand{\bibinfo}[2]{#2}
\providecommand{\BIBentrySTDinterwordspacing}{\spaceskip=0pt\relax}
\providecommand{\BIBentryALTinterwordstretchfactor}{4}
\providecommand{\BIBentryALTinterwordspacing}{\spaceskip=\fontdimen2\font plus
\BIBentryALTinterwordstretchfactor\fontdimen3\font minus
  \fontdimen4\font\relax}
\providecommand{\BIBforeignlanguage}[2]{{%
\expandafter\ifx\csname l@#1\endcsname\relax
\typeout{** WARNING: IEEEtran.bst: No hyphenation pattern has been}%
\typeout{** loaded for the language `#1'. Using the pattern for}%
\typeout{** the default language instead.}%
\else
\language=\csname l@#1\endcsname
\fi
#2}}
\providecommand{\BIBdecl}{\relax}
\BIBdecl

\bibitem{EuropeanUnion2009a}
{European Union}, ``{Directive of 2009/72/EC of the European Parliament and of
  the Council of 13 July 2009 Concerning Common Rules for the Internal Market
  in Electricity and Repealing Directive 2003/54/EC},'' \emph{Official Journal
  of the European Union}, vol. L211, no. August, pp. 55 -- 93, 2009.

\bibitem{Quinn2009}
\BIBentryALTinterwordspacing
E.~L. Quinn, ``{Privacy and the New Energy Infrastructure},'' Tech. Rep.~09,
  2009. [Online]. Available: \url{http://papers.ssrn.com/abstract=1370731}
\BIBentrySTDinterwordspacing

\bibitem{McDaniel2009}
P.~McDaniel and S.~McLaughlin, ``{Security and privacy challenges in the smart
  grid},'' \emph{IEEE Security and Privacy}, vol.~7, no.~3, pp. 75--77, 2009.

\bibitem{Hart1989}
G.~W. Hart, ``{Residential energy monitoring and computerized surveillance via
  utility power flows},'' \emph{IEEE Technology and Society Magazine}, vol.~8,
  no.~2, pp. 12--16, 1989.

\bibitem{Altrabalsi2014}
H.~Altrabalsi, J.~Liao, L.~Stankovic, and V.~Stankovic, ``{A low-complexity
  energy disaggregation method: Performance and robustness},'' in \emph{IEEE
  Symposium on Computational Intelligence Applications in Smart Grid (CIASG)},
  2014.

\bibitem{Guanchen2015}
Z.~Guanchen, G.~Wang, H.~Farhangi, and A.~Palizban, ``{Residential electric
  load disaggregation for low-frequency utility applications},'' in \emph{IEEE
  Power \& Energy Society General Meeting}, 2015.

\bibitem{Molina-Markham2010}
A.~Molina-Markham, P.~Shenoy, K.~Fu, E.~Cecchet, and D.~Irwin, ``{Private
  memoirs of a smart meter},'' in \emph{Proceedings of the 2nd ACM Workshop on
  Embedded Sensing Systems for Energy-Efficiency in Building}.\hskip 1em plus
  0.5em minus 0.4em\relax Zurich, Switzerland: ACM, 2010, pp. 61--66.

\bibitem{Lisovich2010}
M.~A. Lisovich, D.~K. Mulligan, and S.~B. Wicker, ``{Inferring personal
  information from demand-response systems},'' \emph{IEEE Security and
  Privacy}, vol.~8, no.~1, pp. 11--20, 2010.

\bibitem{Greveler2012}
U.~Greveler, B.~Justus, and D.~Loehr, ``{Multimedia content identification
  through smart meter power usage profiles},'' \emph{Computers, Privacy and
  Data Protection}, vol.~1, 2012.

\bibitem{McLaughlin2010}
S.~McLaughlin, D.~Podkuiko, S.~Miadzvezhanka, A.~Delozier, and P.~McDaniel,
  ``{Multi-vendor penetration testing in the advanced metering
  infrastructure},'' in \emph{Proceedings of the 26th Annual Computer Security
  Applications Conference}, Austin, Texas, USA, 2010, pp. 107--116.

\bibitem{Hoenkamp2011}
R.~Hoenkamp, G.~B. Huitema, and A.~J.~C. de~Moor-van Vugt, ``{The neglected
  consumer: the case of the smart meter rollout in the Netherlands},''
  \emph{Renewable Energy Law and Policy Review}, no.~4, pp. 269--282, 2011.

\bibitem{Acs2011}
G.~Acs and C.~Castelluccia, ``{I have a DREAM! (DiffeRentially PrivatE smart
  Metering)},'' \emph{The 13th Information Hiding Conference (IH)}, pp.
  118--132, 2011.

\bibitem{Dong2015}
C.~Dong, S.~Kalra, D.~Irwin, P.~Shenoy, and J.~Albrecht, ``{Preventing
  occupancy detection from smart meters},'' \emph{IEEE Transactions on Smart
  Grid}, vol.~6, no.~5, pp. 2426--2434, 2015.

\bibitem{Kalogridis2010}
G.~Kalogridis, C.~Efthymiou, S.~Z. Denic, T.~A. Lewis, and R.~Cepeda,
  ``{Privacy for Smart Meters: Towards Undetectable Appliance Load
  Signatures},'' in \emph{First IEEE International Conference on Smart Grid
  Communications (SmartGridComm)}, 2010, pp. 232--237.

\bibitem{McLaughlin2011}
S.~Mclaughlin, P.~McDaniel, and W.~Aiello, ``{Protecting consumer privacy from
  electric load monitoring},'' in \emph{Proceedings of the 18th ACM Conference
  Computer and Communications Security (CCS '11)}, Chicago, Illinois, USA,
  2011, pp. 87--98.

\bibitem{Yang2012}
W.~Yang, N.~Li, Y.~Qi, W.~Qardaji, S.~McLaughlin, and P.~McDaniel,
  ``{Minimizing private data disclosures in the smart grid},'' in
  \emph{Proceedings of the 2012 ACM Conference on Computer and Communications
  Security (CCS '12)}.\hskip 1em plus 0.5em minus 0.4em\relax Raleigh, North
  Carolina, USA: ACM, 2012, p. 415.

\bibitem{Giaconi2016}
G.~Giaconi and D.~Gunduz, ``{Smart meter privacy with renewable energy and a
  finite capacity battery},'' in \emph{IEEE Workshop on Signal Processing
  Advances in Wireless Communications, SPAWC}, vol. 2016-August, Edinburgh, UK,
  2016.

\bibitem{Li2015}
S.~Li, A.~Khisti, and A.~Mahajan, ``{Structure of optimal privacy-preserving
  policies in smart-metered systems with a rechargeable battery},'' in
  \emph{IEEE Workshop on Signal Processing Advances in Wireless Communications,
  SPAWC}, 2015, pp. 375--379.

\bibitem{Tan2013}
O.~Tan, D.~Gunduz, and H.~V. Poor, ``{Increasing smart meter privacy through
  energy harvesting and storage devices},'' \emph{IEEE Journal on Selected
  Areas in Communications}, vol.~31, no.~7, pp. 1331--1341, 2013.

\bibitem{Agency2015}
\BIBentryALTinterwordspacing
{IRENA}, ``{Battery Storage for Renewables: Market Status and Technology
  Outlook},'' Tech. Rep., 2015. [Online]. Available:
  \url{http://www.irena.org/DocumentDownloads/Publications/IRENA\_Battery\_Storage\_report\_2015.pdf}
\BIBentrySTDinterwordspacing

\bibitem{Khakimova2015}
A.~Khakimova, A.~Shamshimova, D.~Sharipova, A.~Kusatayeva, V.~Ten, A.~Bemporad,
  Y.~Familiant, A.~Shintemirov, and M.~Rubagotti, ``{Hybrid model predictive
  control for optimal energy management of a smart house},'' in \emph{IEEE 15th
  International Conference on Environment and Electrical Engineering (EEEIC)},
  2015, pp. 513--518.

\bibitem{Haiming2015}
W.~Haiming, M.~Ke, D.~Zhao~Yang, X.~Zhao, L.~Fengji, and W.~Kit~Po,
  ``{Efficient real-time residential energy management through MILP based
  rolling horizon optimization},'' in \emph{IEEE Power \& Energy Society
  General Meeting}, 2015.

\bibitem{Galka2005}
A.~Galka, T.~Ozaki, and O.~Yamashita, ``{A new approach to mutual information
  between pairs of times series},'' in \emph{International Symposium on
  Nonlinear Theory and its Applications (NOLTA2005)}, Bruges, Belgium, 2005,
  pp. 626--629.

\bibitem{Manning2008}
C.~D. Manning, P.~Raghavan, and H.~Schutze, \emph{{An Introduction to
  Information Retrieval}}.\hskip 1em plus 0.5em minus 0.4em\relax Cambridge
  University Press, 2009.

\bibitem{Lofberg2012}
J.~L{\"{o}}fberg, ``{Oops! I cannot do it again: Testing for recursive
  feasibility in MPC},'' \emph{Automatica}, vol.~48, no.~3, pp. 550--555, 2012.

\bibitem{Lofberg2004}
J.~Lofberg, ``{YALMIP : a toolbox for modeling and optimization in MATLAB},''
  in \emph{IEEE International Conference on Computer Aided Control Systems
  Design}, 2004, pp. 284--289.

\bibitem{Pflugradt2013}
N.~Pflugradt and W.~Schufft, ``{Analysing low-voltage grids using a behaviour
  based load profile generator},'' \emph{Renewable Energy and Power Quality
  Journal}, vol.~1, no.~11, pp. 1--5, 2013.

\end{thebibliography}

% insert where needed to balance the two columns on the last page with
% biographies
\newpage
\begin{IEEEbiography}
    [{\includegraphics[width=1in,height=1.25in,clip,keepaspectratio]{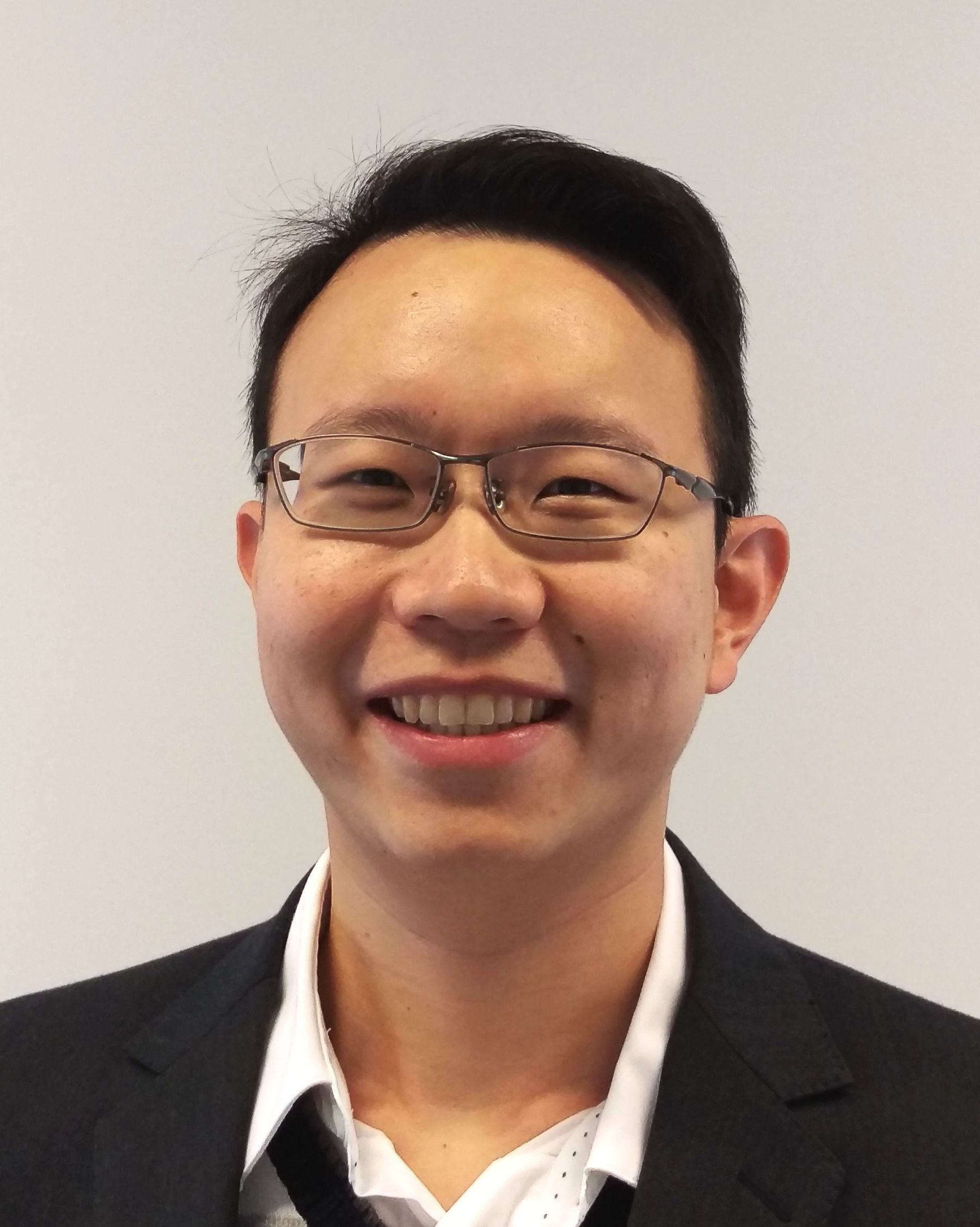}}]{Jun-Xing Chin}
(S'15) received the B.Eng. degree in electronic engineering and the B.BA degree in general business management from the Hong Kong University of Science and Technology in 2010, and the M.Sc. degree in electrical technology for sustainable and renewable energy systems from the University of Nottingham, the United Kingdom in 2011. From 2011 to 2015, he was an electrical engineer at the engineering consultancy firm Ove Arup and Partners H.K. Ltd. in Hong Kong. Currently, he is working towards his Ph.D. degree at the Power Systems Laboratory at ETH Zurich, with research emphasis on consumer privacy for smart metering systems, and short-term load forecasting. 
\end{IEEEbiography}
\vspace{-3cm}
\begin{IEEEbiography}
    [{\includegraphics[width=1in,height=1.25in,clip,keepaspectratio]{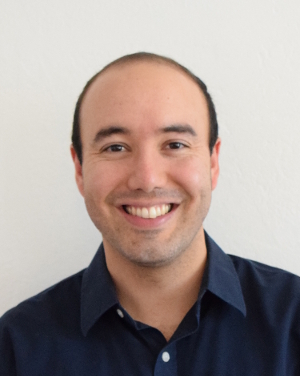}}]{Tomas Tinoco De Rubira}
	received the B.S. degree in Electrical and Computer Engineering from the University of California Berkeley in 2008, and the M.S. and Ph.D. degrees in Electrical Engineering from Stanford University in 2011 and 2015, respectively. He is currently a Postdoctoral Research Fellow at the Power Systems Laboratory of ETH Zurich. His research consists on developing and applying mathematical optimization techniques for solving challenging problems, such as enabling large-scale integration of renewable energy in electric power grids. His employment experience includes Aurora Solar Inc. and the Electric Power Research Institute.
\end{IEEEbiography}
\vspace{-3cm}
\begin{IEEEbiography}
    [{\includegraphics[width=1in,height=1.25in,clip,keepaspectratio]{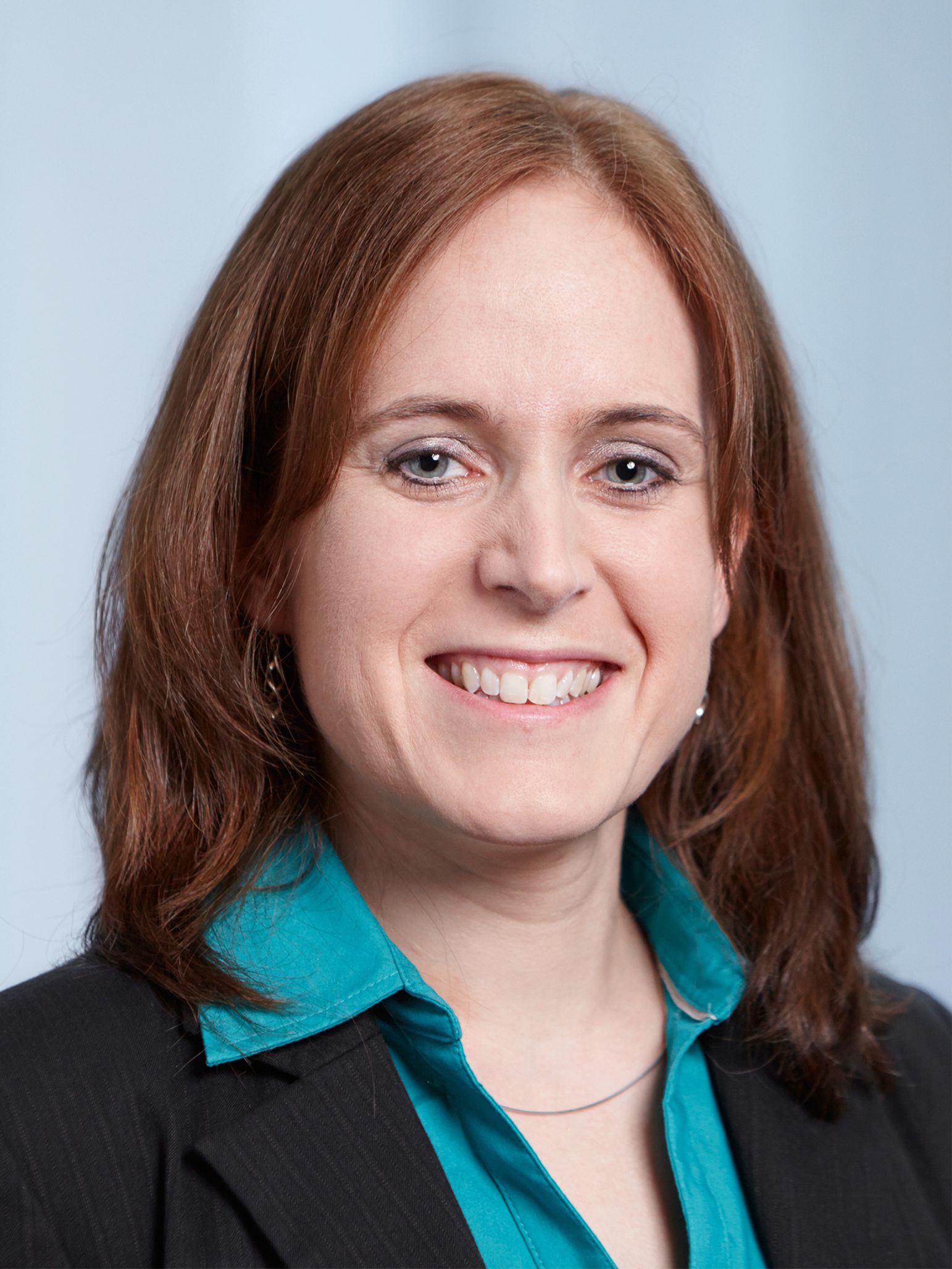}}]{Gabriela Hug}
(S'05, M'08, SM'14) was born in Baden, Switzerland. She received the M.Sc. degree in electrical engineering from the Swiss Federal Institute of Technology in Zurich (ETH Zurich) in 2004 and the Ph.D. degree from the same institution in 2008. After her Ph.D., she worked in the Special Studies Group at Hydro One in Toronto, Canada. From 2009 to 2015, she was an Assistant Professor at Carnegie Mellon University in Pittsburgh, Pennsylvania. Currently, she is an Associate Professor in the Power Systems Laboratory at ETH Zurich. Her research is dedicated to the control and optimization of electric power systems.
\end{IEEEbiography}
\vspace{9cm}
~
\end{document}